\documentclass[useAMS,usenatbib]{mn2e}

\usepackage{times,graphicx,amssymb,natbib}

\newcommand{\degrees}{^{\circ}}
\newcommand{\msol}{M_{\rm \odot}}
\newcommand{\mjup}{M_{\rm Jup}}

\title[Circumbinary Habitable Zones]{Assessing Circumbinary Habitable Zones using Latitudinal Energy Balance Modelling}
\author[D. Forgan]{Duncan Forgan $^{1}$\thanks{E-mail: dhf@roe.ac.uk}
  \\ $^{1}$Scottish Universities Physics Alliance (SUPA), Institute
  for Astronomy, University of Edinburgh, Blackford Hill, Edinburgh,
  EH9 3HJ, Scotland, UK\\}
\begin{document}

\date{Accepted}

\pagerange{\pageref{firstpage}--\pageref{lastpage}} \pubyear{}

\maketitle

\label{firstpage}

\begin{abstract}


\noindent Previous attempts to describe circumbinary habitable zones have been concerned with the spatial extent of the zone, calculated analytically according to the combined radiation field of both stars.  By contrast to these ``spatial HZs'', we present a numerical analysis of the ``orbital HZ'', a habitable zone defined as a function of planet orbital elements. This orbital HZ is better equipped to handle (for example) eccentric planet orbits, and is more directly connected to the data returned by exoplanet observations.  

Producing an orbital HZ requires a large number of climate simulations to be run to investigate the parameter space - we achieve this using Latitudinal Energy Balance Models (LEBMs), which handle the insolation of the planet by both stars (including mutual eclipses), as well as the planetary atmosphere's ability to absorb, transfer and lose heat. 

We present orbital HZs for several known circumbinary planetary systems: Kepler-16, Kepler-34, Kepler-35, Kepler-47 and PH-1.  Generally, the orbital HZs at zero eccentricity are consistent with spatial HZs derived by other authors, although we detect some signatures of variability that coincide with resonances between the binary and planet orbital periods.  We confirm that Earthlike planets around Kepler-47 with Kepler-47c's orbital parameters could possess liquid water, despite current uncertainties regarding its eccentricity.  Kepler-16b is found to be outside the habitable zone, as well as the other circumbinary planets investigated.

\end{abstract}

\begin{keywords}
astrobiology, planets and satellites: general, methods:numerical
\end{keywords}

\section{Introduction}


\noindent The Habitable Zone (HZ) is a useful conceptual tool in investigating the general habitability of planetary systems.  It is usually defined as the region surrounding a star which, if a terrestrial planet of Earth mass and similar atmospheric composition were to reside within it, the water upon the planet's surface would remain liquid \citep{Huang1959,Hart_HZ}.  The boundaries of the HZ are subsequently defined by the properties of the host star: the outer boundary of the HZ is typically governed by the rate at which $CO_2$ clouds maintain a sufficiently strong greenhouse effect, and the inner edge of the HZ is controlled by the rate of water loss via hydrogen escape and hydrolysis.  The HZ is sensitive to the spectrum of the source of insolation - in particular, how strongly the source emits in the infrared (IR).  As a result, the inner and outer boundaries of the HZ are a function of the effective temperature of the star.

While the majority of the literature utilising the HZ concept has relied on the seminal atmospheric radiative transfer calculations of \citet{Kasting_et_al_93}, and subsequent parametrisations (e.g. \citealt{Underwood2003,Selsis2007,Kaltenegger2011}), it should be noted that \citet{Kopparapu2013} have since returned to these calculations, updating the atmospheric absorption models and extending the range of stellar effective temperatures calculated.  This has the effect of moving the conservative HZ boundaries for the Solar System outwards slightly.  

For a single star, the HZ boundary conditions are spherically symmetric, and as a result, the single-star HZ is a circular annulus.  Therefore, planets of Earth mass and atmospheric pressure/composition on circular orbits within the HZ are expected to possess liquid water, and hence be potentially habitable.  If the planet's orbit is elliptical, but impinges upon the HZ, then it can be habitable depending on the average flux received by the planet over the orbit, or equivalently how long it spends within the habitable zone \citep{Williams2002, Kane2012,Kane2012c}.  

Since the first detection of an exoplanet orbiting a main sequence star \citep{Mayor1995}, the science of exoplanet detection has quickly revealed a large number of exoplanets, with several residing in their local (single-star) HZs, e.g. Kepler-22b \citep{Borucki2012c}, Kepler-62f \citep{Borucki2013}, or the three planets Gliese 667Cc, 667Ce and 667Cf, which occupy the same HZ \citep{Anglada-Escude2013}.  However, these two Kepler planets possess radii 1.4 to 2.4 times larger than that of the Earth, and the Gliese 667C planets have masses greater than two Earth masses.  Combined with the current ignorance as to their atmospheric composition, it is unclear if these objects are themselves habitable\footnote{It should also be noted that liquid water is considered one of the primary necessary conditions for habitability, but it is unlikely to be a sufficient condition.  Extrapolating astrobiological data from a single data point (the Earth) is demonstrably difficult (cf \citealt{Spiegel2012})}.  Equally, these objects could possess exomoons which may themselves be habitable \citep{Forgan_moon1}, and the detection of Earth-mass exomoons is now possible with current observations \citep{Kipping2013a}.  

The growing exoplanet population continues to challenge our preconceptions of what can constitute a stable, potentially habitable planetary system.  Binary star systems are among the most recent of these exotic systems to be discovered.  In S-type binary systems, such as Alpha Centauri, the binary typically has a sufficiently large semimajor axis (of order 10 - 50 AU) that stable planetary orbits exist around either of the two stars.  It has been established by numerical simulation \citep{Wiegert1997,Quintana2002,Quintana2007} that S-type binary systems can form planets in habitable regions around one or both stars.  In this scenario, provided that the distance between the two stars remains sufficiently large, approximating the system's habitable zone with two single star HZs placed around the binary components is usually acceptable.  If this is not the case, e.g. if the binary eccentricity is large, then more detailed calculations are required (e.g. \citealt{Forgan2012,Eggl2012,Kaltenegger2013}).   

In the P-type ``circumbinary'' systems, the stars orbit sufficiently closely that the planet orbits the system's centre of mass, and the single-star approximation clearly fails.  \citet{Kane2013} produced analytical calculations which approximate the aggregrate stellar flux as a blackbody function, with a peak wavelength equal to that found by adding the flux from both stars.  Applying Wien's Law yields a combined effective temperature, which can then be used in conjunction with the bolometric flux to calculate HZ boundaries using the single star HZ prescriptions \citep{Kasting_et_al_93, Underwood2003}.  In a similar vein, \citet{Haghighipour2013} also use the single-star HZ prescriptions, weighting the flux received from each star at a given location according to its effective temperature, and searching for the points where the weighted flux equals the flux received from a 1 $\msol$ star at the inner and outer boundaries.  Both methods produce similar calculations for the combined habitable zones, which can deviate strongly from the circular annuli depending on the binary mass ratio and orbital elements.

While habitable zones can be defined spatially as described above, they can also be defined by the set of allowed planetary orbital elements that permit liquid water on their surface.  Instead of analytically calculating what we might call ``the spatial HZ'', and measuring the time that planets spend within the zone, we can attack the problem numerically, by evolving the climates of many planets on a multidimensional grid of orbital elements, mapping out an `` orbital HZ'' in this parameter space.  While the orbital HZ may not supply the same level of theoretical insight as a spatial HZ, it does possess two advantages:

\begin{enumerate}
\item The spatial HZs in multiple star systems are complex and time-dependent, and hence the time required to calculate a planet's habitability using the spatial HZ increases quickly as the number of stars in the system increases.  Conversely, numerical simulations that produce an orbital HZ typically demonstrate a weaker scaling of compute time with star number.
\item Simulations such as those used to generate the orbital HZ can incorporate the effect of stellar eclipses easily.  For analytic calculations, some parametrisations are available (cf \citealt{Heller2012}) but this has not yet been done for P-type binary systems.
\item Exoplanet observations produce orbital parameters as output. As such, astrobiologists adopting an orbital HZ will have a more immediate and profound grasp on the habitability of an exoplanet than they might obtain by constructing a spatial HZ as an intermediate step.
\end{enumerate}

\noindent Identifying orbital HZs requires running a large number of individual simulations.  The climate model used must therefore be fast, robust, and reliable.  Latitudinal energy balance models (LEBMs) are well suited to this task \citep{North1981, Williams1997a,Williams2002, Spiegel_et_al_08, Spiegel2009,Dressing2010,Spiegel2010,Forgan2012, Forgan_moon1,Vladilo2013}.  By splitting the planet into latitudinal strips, making some simplifying assumptions about atmospheric stratification, and the spectral energy distribution of the incoming radiation, LEBMs require very little CPU time to complete a climate simulation that faithfully reproduces climates on Earthlike planets (see e.g. \citealt{Spiegel_et_al_08} or \citealt{Vladilo2013} for examples of tests).

In this work, we use LEBMs to assess the orbital HZs in several known circumbinary planetary systems: Kepler-16, Kepler-34, Kepler-35, Kepler-47 and PH1. We investigate the HZ as a function of planet semimajor axis $a_p$ and planet eccentricity $e_p$, and compare the LEBM calculations to analytic calculations of circumbinary habitable zones.

In section \ref{sec:Method} we describe the construction of the LEBM and the initial conditions used; in section \ref{sec:Results} we display the resulting orbital HZs produced using the LEBMs for the above circumbinary systems.  In section \ref{sec:Discussion}, we investigate the dependence of the circumbinary HZs on the orbital parameters of the binary, and suggest routes for future improvement, and in section \ref{sec:Conclusions} we summarise the work.

\section{Method }\label{sec:Method}

\subsection{Latitudinal Energy Balance Models}

\noindent The LEBM is a one dimensional diffusion equation of surface temperature:

\begin{equation} 
C \frac{\partial T(x,t)}{\partial t} - \frac{\partial }{\partial x}\left(D(1 - x^2)\frac{\partial T(x,t)}{\partial x}\right) = S(1-A(T)) - I(T). 
\end{equation}

\noindent Rather than using the latitude, $\lambda$, directly, the variable $x= \sin \lambda$ is used instead for reasons of computational expediency (the $(1-x^2)$ term being a geometric factor arising from the spherical geometry of the problem).  This equation is evolved with the boundary condition $\frac{dT}{dx}=0$ at the poles (where $\lambda=[-90,90] \degrees$).   

$T(x,t)$ is the surface temperature, $C$ is the effective heat capacity of the atmosphere, $D$ is a diffusion coefficient that determines the efficiency of heat redistribution across latitudes, $S$ is the  insolation flux, $I$ is the IR cooling and $A$ is the albedo.  In the above equation, $C$, $S$, $I$ and $A$ are functions of $x$ (either explicitly, as $S$ is, or implicitly through $T$).   

The diffusion constant $D$ is defined such that a planet at 1 au around a star of $1 \msol$, with rotation period of 1 day will reproduce the average temperature profile measured on Earth (see e.g. \citealt{Spiegel_et_al_08}).  Planets that rapidly rotate experience inhibited latitudinal heat transport, due to Coriolis effects, resulting in a $D \propto \omega_d^{-2}$ scaling, where $\omega_d$ is the rotational angular velocity of the planet (see \citealt{Farrell1990}).  We therefore use:

\begin{equation} 
D=5.394 \times 10^2 \left(\frac{\omega_d}{\omega_{d,\oplus}}\right)^{-2},\label{eq:D}
\end{equation}

\noindent where $\omega_{d,\oplus}$ is the rotational angular velocity of the Earth.   This expression is certainly too simple to describe the full effects of rotation, as more detailed global circulation modelling indicates \citep{DelGenio1993,DelGenio1996}.  A more rigorous expression would include the effects of atmospheric pressure and mean molecular weight (e.g. \citealt{Williams1997a}, but see also \citealt{Vladilo2013}'s attempts to introduce a latitudinal dependence to $D$ to mimic the Hadley convective cells on Earth).  

\noindent As in \citet{Forgan2012} and \citet{Forgan_moon1}, we solve the diffusion equation using an explicit forward time, centre space finite difference algorithm.  A global timestep was adopted, with constraint

\begin{equation}
\delta t < \frac{\left(\Delta x\right)^2C}{2D(1-x^2)}.  
\end{equation}

As the system is longitudinally averaged, a key assumption of the model (and its inputs) is that the planet rotates sufficiently quickly relative to its orbital period.  We adopt the same input expressions for the atmospheric heat capacity, albedo, insolation and atmospheric cooling as was done in \citet{Forgan2012}, which we summarise here.

The atmospheric heat capacity depends on what fraction of the planet's surface is ocean, $f_{ocean}$, what fraction is land $f_{land}=1.0-f_{ocean}$, and what fraction of the ocean is frozen $f_{ice}$:

\begin{equation} 
C = f_{land}C_{land} + f_{ocean}\left((1-f_{ice})C_{ocean} + f_{ice} C_{ice}\right). 
\end{equation}

\noindent The heat capacities of land, ocean and ice covered areas are

\begin{equation} C_{land} = 5.25 \times 10^9 $ erg cm$^{-2}$ K$^{-1}\end{equation}
\begin{equation} C_{ocean} = 40.0C_{land}\end{equation}
\begin{equation} C_{ice} = \left\{
\begin{array}{l l }
9.2C_{land} & \quad \mbox{263 K $< T <$ 273 K} \\
2C_{land} & \quad \mbox{$T<263$ K}. \\
\end{array} \right. \end{equation}

\noindent The infrared cooling function is 

\begin{equation} 
I(T) = \frac{\sigma_{SB}T^4}{1 +0.75 \tau_{IR}(T)}, 
\end{equation}

\noindent where the optical depth of the atmosphere 

\begin{equation} 
\tau_{IR}(T) = 0.79\left(\frac{T}{273\,\mathrm{K}}\right)^3. 
\end{equation}

\noindent The albedo function is

\begin{equation} 
A(T) = 0.525 - 0.245 \tanh \left[\frac{T-268\,\mathrm K}{5\, \mathrm K} \right]. 
\end{equation}

\noindent As the surface temperature drops and water freezes, the albedo increases rapidly and non-linearly.  This sets up a positive feedback loop that can make the outer HZ extremely sensitive to small perturbations in dynamical or radiative properties.

At any instant, for a single star, the insolation received at a given latitude at an orbital distance $r$ is

\begin{equation}
S = q_0\cos Z \left(\frac{1 AU}{r}\right)^2,
\end{equation}

\noindent where $q_0$ is the bolometric flux received from the star at a distance of 1 AU, and $Z$ is the zenith angle:

\begin{equation} q_0 = 1.36\times 10^6\left(\frac{M}{\msol}\right)^4
  \mathrm{erg \,s^{-1} cm^{-2}} 
  \end{equation}

\begin{equation} \cos Z = \mu = \sin \lambda \sin \delta + \cos
  \lambda \cos \delta \cos h. \end{equation} 

\noindent Here, we assume the luminosity can be determined from main sequence scaling ($\msol$ represents one solar mass).  In this form, the model cannot describe binary systems with post-main sequence components, but in principle it can be updated to do so, provided that the spectra of the stars are well described.

The solar hour angle is $h$, and $\delta$ is the solar declination, which is calculated from the obliquity $\delta_0$ using:

\begin{equation} 
\sin \delta = -\sin \delta_0 \cos(\phi_p-\phi_{peri}-\phi_a), 
\end{equation}

\noindent where $\phi_p$ is the current orbital longitude of the planet, $\phi_{peri}$ is the longitude of periastron, and $\phi_a$ is the longitude of winter solstice, relative to the longitude of periastron.  

As we use diurnally averaged quantities, we must also diurnally average $S$:

\begin{equation} S = q_0 \bar{\mu}. \end{equation}

\noindent We do this by integrating $\mu$ over the sunlit part of the day, i.e. $h=[-H, +H]$, where $H(x)$ is the radian half-day length at a given latitude.  Multiplying by $H/\pi$ (as $H=\pi$ if a latitude is illuminated for a full rotation) gives the total diurnal insolation as

\begin{equation} 
S = q_0 \left(\frac{H}{\pi}\right) \bar{\mu} = \frac{q_0}{\pi} \left(H \sin \lambda \sin \delta + \cos \lambda \cos \delta \sin H\right). 
\end{equation}

\noindent The radian half day length is calculated as

\begin{equation} 
\cos H = -\tan \lambda \tan \delta. 
\end{equation}

\noindent Both stars contribute to the total flux $S$.  We calculate the orbital longitude, solar declination and radian half-day length for both stars, as well as the distance of the planet from both stars.  If one star is eclipsed by the other, then we set its contribution to $S$ to zero.  We ensure that the simulation can accurately model a transit by adding an extra timestep criterion, ensuring that the transit's duration will not be less than ten timesteps.  

\subsection{Determining the Habitable Zone - Classification of Model Outcomes}

\noindent When using LEBMs, it is common to calculate a ``habitability function'' $\xi$ (see \citealt{Spiegel_et_al_08}):

\begin{equation} \xi(\lambda,t) = \left\{
\begin{array}{l l }
1 & \quad \mbox{273 K $< T(\lambda,t) <$ 373 K} \\
0 & \quad \mbox{otherwise}. \\
\end{array} \right. 
\end{equation}

\noindent Strictly, this is a potential habitability function - it simply measures whether a given latitude lies in the temperature range where liquid water may exist.  This paper relies heavily on this function, and discussions of habitability refer specifically to the fraction of surface where liquid water may exist.

We average this function over latitude to calculate the fraction of potentially habitable surface at time $t$:

\begin{equation} 
\xi(t) = 1/2 \int_{-\pi/2}^{\pi/2}\xi(\lambda,t)\cos \lambda \,d\lambda. 
 \end{equation}

\noindent We will use this function to classify the planets we simulate in the following sections.  Once each simulation has settled into a quasi-steady state, we average $\xi$ over the last ten years of the run, and use the mean, $\bar{\xi}$, and its standard deviation $\sigma_\xi$, to classify as follows:

\begin{enumerate}
\item \emph{Habitable Planets} - these planets exhibit $\bar{\xi}>0.1$, and $\sigma_{\xi} < 0.1\bar{\xi}$, i.e. the fluctuation in habitable surface is less than 10\% of the mean.
\item \emph{Hot Planets} - these planets have temperatures above 373 K across all seasons, and are therefore completely uninhabitable ($\bar{\xi} <0.1$).
\item \emph{Snowball Planets} - these planets are completely frozen and are therefore completely uninhabitable ($\bar{\xi}<0.1$).
\item \emph{Transient Planets} - these planets possess a time-averaged $\bar{\xi}>0.1$, but $\sigma_{\xi} > 0.1\bar{\xi}$, i.e. the fluctuation in habitable surface is greater than 10\% of the mean.
\end{enumerate}

\begin{figure}
\begin{center}
\includegraphics[scale = 0.5]{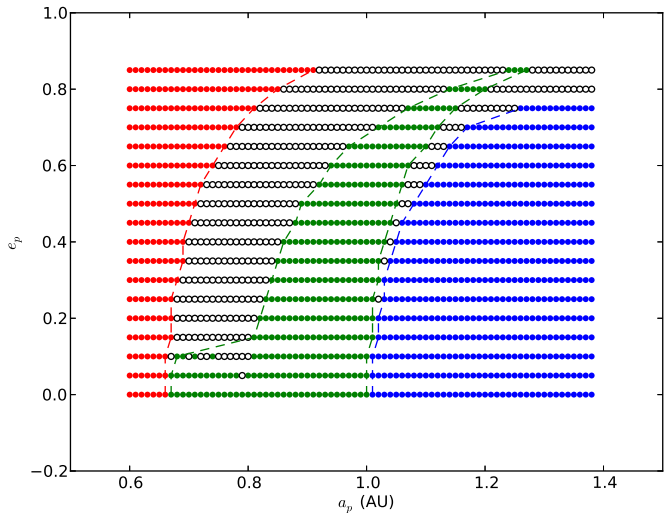} 
\caption{The habitable zone for an Earth-like planet around a Sun-like star, as calculated from a LEBM using the classification system outlined above. We plot the results for each simulation according to the planet's semimajor axis (x-axis) and the planet's eccentricity (y-axis), and the colour of the point indicates its outcome.  Red points are hot planets with no habitable surface; blue points are cold planets with no habitable surface; green points represent warm planets with at least ten percent of the surface habitable and low seasonal fluctuations; white circles represent warm planets with high seasonal fluctuations. The dashed lines indicate boundaries between classifications.  The green dashed lines indicate the conventional habitable zone.  \label{fig:sunhighres}}
\end{center}
\end{figure}

\noindent Figure \ref{fig:sunhighres} shows the single-star habitable zone for the Solar system as it would be classified by the above taxonomy.  Note the extension of the habitable zone (as described by the green points) to low semimajor axis at low eccentricity.  This is a symptom of only requiring $\bar{\xi}>0.1$ for habitability.  As the seasonal variations in climate around low eccentricity planets are relatively low, this allows a planet with a fairly inhospitable surface to maintain small habitable regions at the poles which do not vary greatly in extent.

For comparison, the Earth's parameters exhibits $\bar{\xi} \sim 0.85$. This is much higher than the value required to classify a planet as habitable, and it might be suggested that requiring $\bar{\xi} >0.1$ is not particularly demanding.  The habitable zones we delineate here are quite generous, and planets at the edges of the zone will be largely inhospitable, but will still possess regions that remain habitable throughout the season, and as such sufficient to maintain a modest but limited biosphere.

\subsection{Initial Conditions}

\noindent Unless otherwise stated, the planets simulated are assumed to be Earthlike.  The diurnal period is set equal to the Earth's, the obliquity is set to $23.5\degrees$, and the surface ocean fraction $f_{ocean}$ is set to 0.7.  We fix these parameters for expediency, but we should note that these parameters have their own effects on habitability.  Increasing the rotation rate can suppress the latitudinal transport of heat \citep{Farrell1990}. Planets with low surface ocean fractions will experience stronger seasonal temperature variations \citep{Abe2011,Forgan2012} which would have obvious consequences for the classification system used in this paper.  Planets with larger obliquity appear to resist the ``snowball'' transition to a completely frozen state, even when rapid rotation would otherwise encourage it \citep{Spiegel2009}.

The planets orbit in the binary plane, around the centre of mass of the binary system (with the exception of comparison simulations run without the secondary).  The simulations begin at the northern winter solstice, which is assumed to occur at an orbital longitude of $0 \degrees$.  In the case of eccentric orbits, this is also the longitude of periastron\footnote{Simulations were carried out where the longitude of periastron was varied.  As the habitability calculations average over many orbits, the effect of changing the initial phase is minimal}.  The planets' initial temperature was set to 288 K at all latitudes.  Each simulation is run for a sufficient length of time that the planet's temperature profile reaches a periodic, steady state, such that the habitability classifications described earlier can be made.  Table \ref{tab:params} lists the input parameters for all binary systems studied in this paper.

\begin{table}
\centering
\begin{minipage}{140mm}
\caption{Parameters used in this work to describe each binary system.\label{tab:params}}
\begin{tabular}{c | cccc}
\hline
\hline
Name & $M_1 \, (\msol)$ & $M_2 \, (\msol)$ & $a_{\rm bin} \, (AU)$ & $e_{\rm bin}$ \\
\hline
Kepler-16 & 0.6897 & 0.2026 & 0.224 & 0.15944 \\
Kepler-34 & 1.0479 & 1.0208 & 0.224 & 0.52087 \\
Kepler-35 & 0.8877 & 0.8094 & 0.176 & 0.1418 \\
Kepler-47 & 1.043 & 0.362 & 0.0836 & 0.0234 \\
PH1 & 1.384 & 0.386 & 0.144 & 0.0 \\
\hline
\hline
\end{tabular}
\end{minipage}
\end{table}

\section{Results}\label{sec:Results}

\subsection{Kepler-16}

\begin{figure*}
\begin{center}$\begin{array}{cc}
\includegraphics[scale = 0.5]{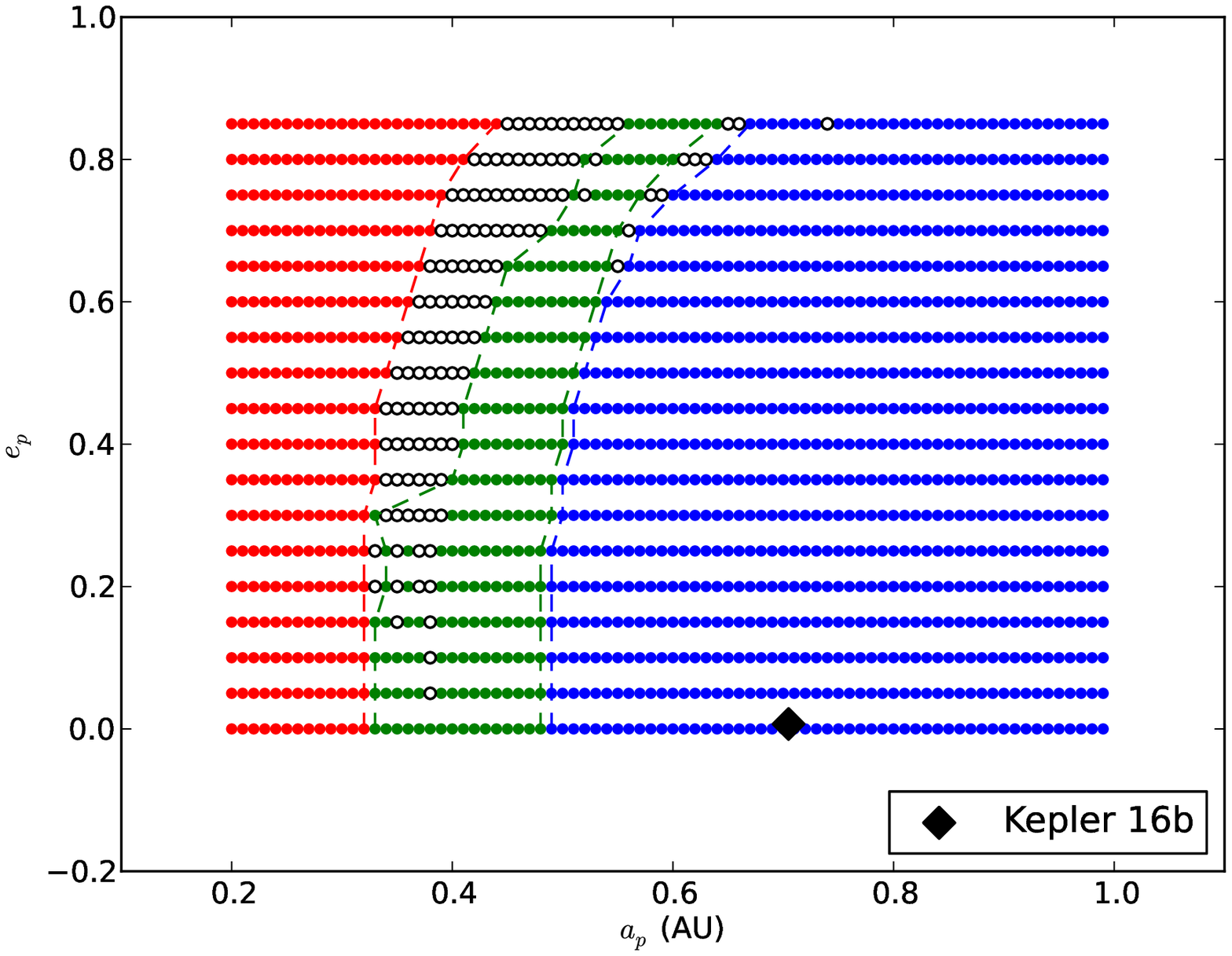}  &
\includegraphics[scale=0.5]{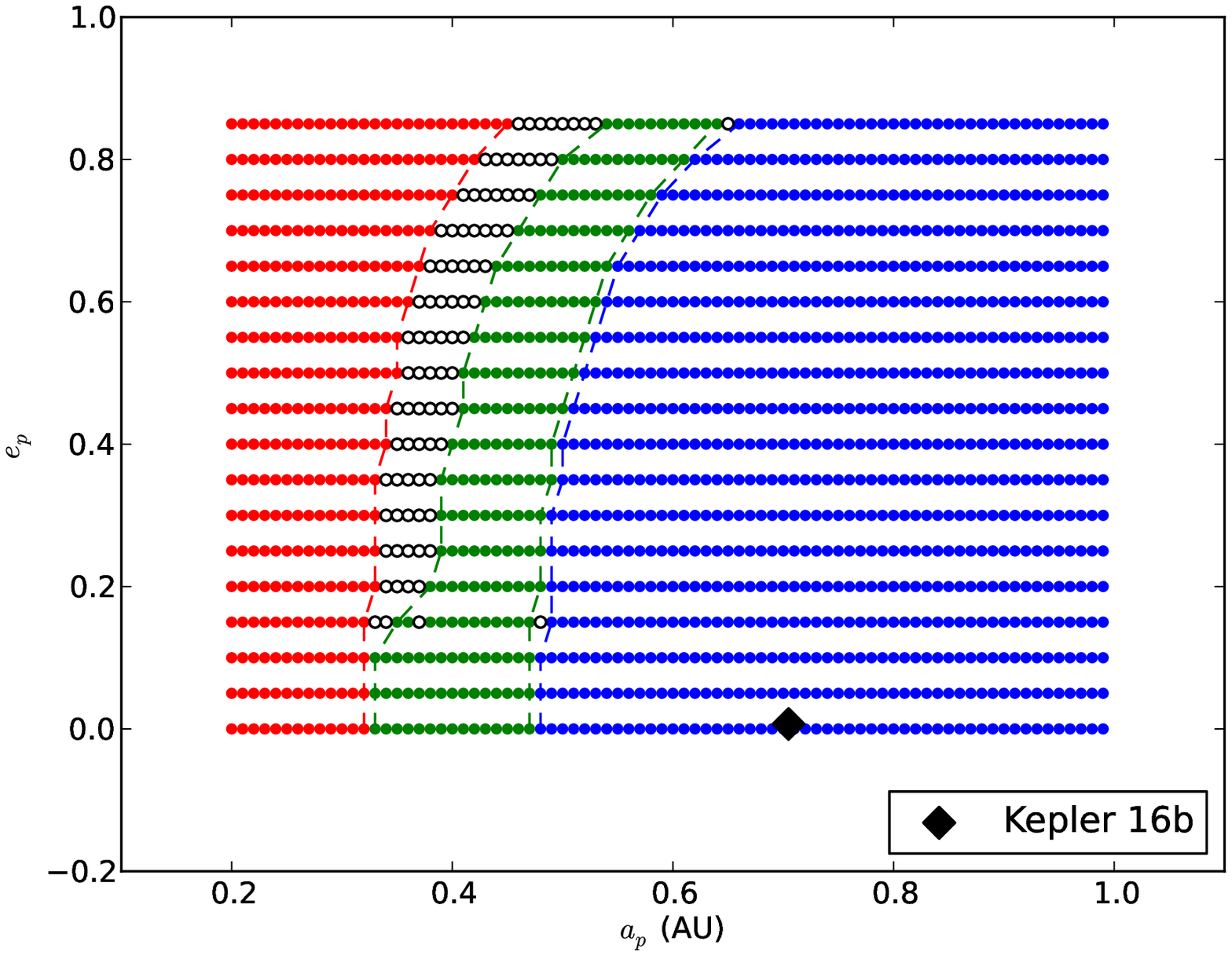} \\
\end{array}$
\caption{Left: The habitable zone for an Earth-like planet in the Kepler-16 binary system; Right: the habitable zone for an Earth-like planet in the Kepler-16 system with the secondary removed.  The colour of points represents the classification of each simulation run.  As before, red points are hot planets with no habitable surface, blue points are cold planets with no habitable surface, green points are warm planets with at least ten percent of the surface habitable and low seasonal fluctuations, and white circles are warm moons with high seasonal fluctuations.  Again, dashed lines denote boundaries between different classifications.
\label{fig:kepler16ab}}
\end{center}
\end{figure*}

\noindent Kepler-16 was the first circumbinary planetary system to be discovered during the Kepler mission \citep{Doyle2011}.  Kepler-16b, with mass 0.3 $\mjup$, orbits the binary with a period of 229 days, while the binary orbital period is 41 days.  The left panel of Figure \ref{fig:kepler16ab} shows the habitable zone in $a_p-e_p$ space for the Kepler-16 binary system.  For comparison, the right hand panel of the same figure shows the equivalent habitable zone in the absence of the secondary star.  Given the large mass difference between the primary and secondary, it is not surprising that HZs produced with and without the secondary are so similar.

What is more interesting is the switch in habitability classification for some parameters from habitable to transient as the secondary star is added.  This might be expected for planets near the inner HZ edge on the right panel of Figure \ref{fig:kepler16ab} - the extra insolation from the second star, coupled with eclipses of the primary by the secondary, can produce temperature variations of sufficient strength to periodically push large fractions of the planet's surface above 373K.  Indeed, this change is strongest at around 0.35 AU, which corresponds to a 2:1 resonance between the planet and binary orbital periods.

However, we also see this reclassification at the outer HZ edge (approximately $a_p=0.6 AU$) at eccentricities above $e_p=0.6$, which is more surprising.  The secondary insolation at this distance from the binary is less than a few percent of the primary insolation.  So how is the surface temperature so strongly affected? At 0.6 AU, planets with eccentricities greater than 0.6 will have periastra located inside the binary's orbit.  These very close approaches to the secondary will produce climate variations that ensure the planet is classified as transient.

The orbital HZ constructed here would suggest that Kepler-16b, which has a close to circular orbit at 0.7 AU,  is too cold to be in the habitable zone, as confirmed by \citet{Kane2013} and \citet{Haghighipour2013}. It has been suggested that Kepler-16b could host a habitable Earth-mass captured planet in a satellite or Trojan orbit \citep{Quarles2012}, but this possibility is outside the scope of this work. 

\subsection{Kepler-34}

\begin{figure*}
\begin{center}$\begin{array}{cc}
\includegraphics[scale = 0.5]{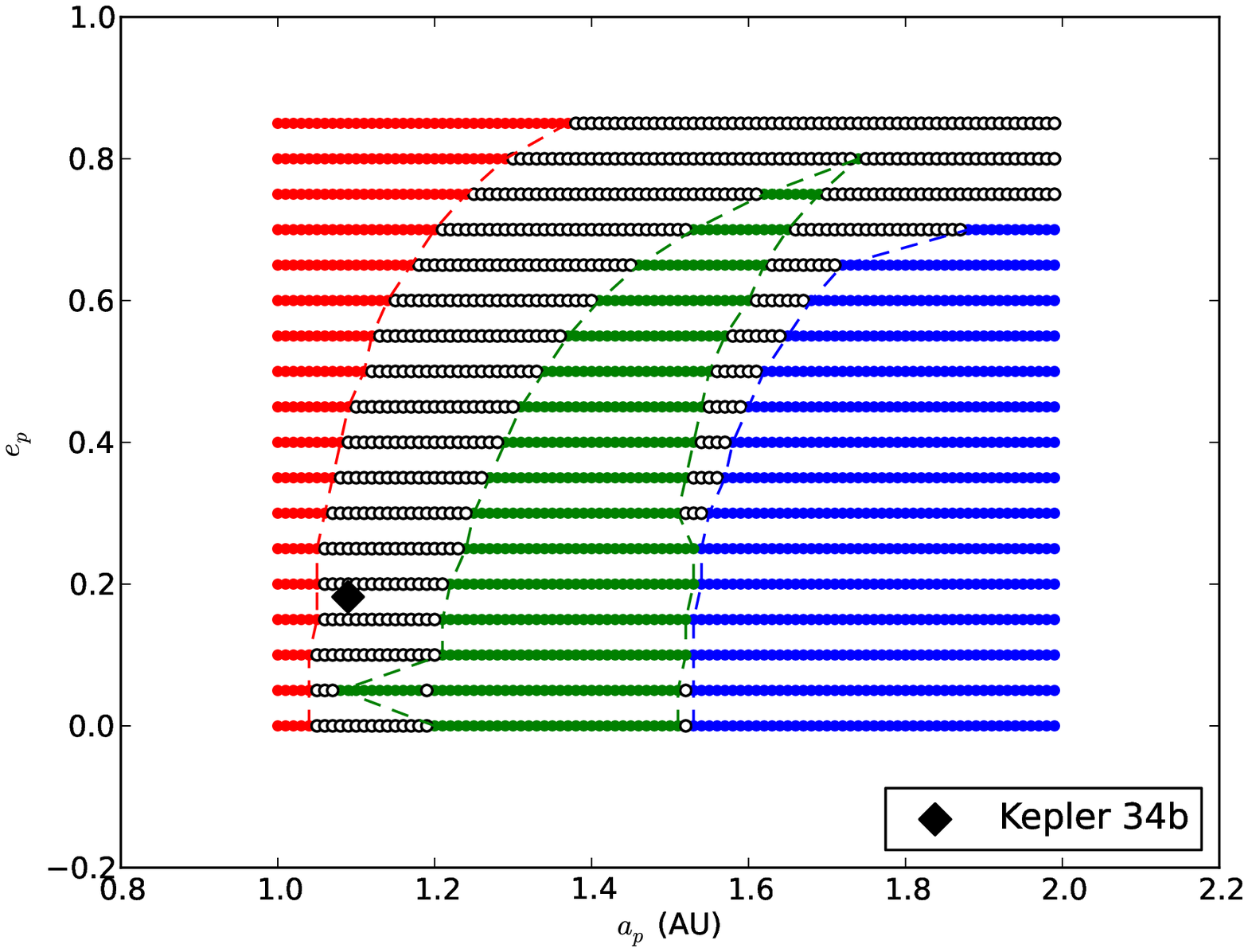}  &
\includegraphics[scale=0.5]{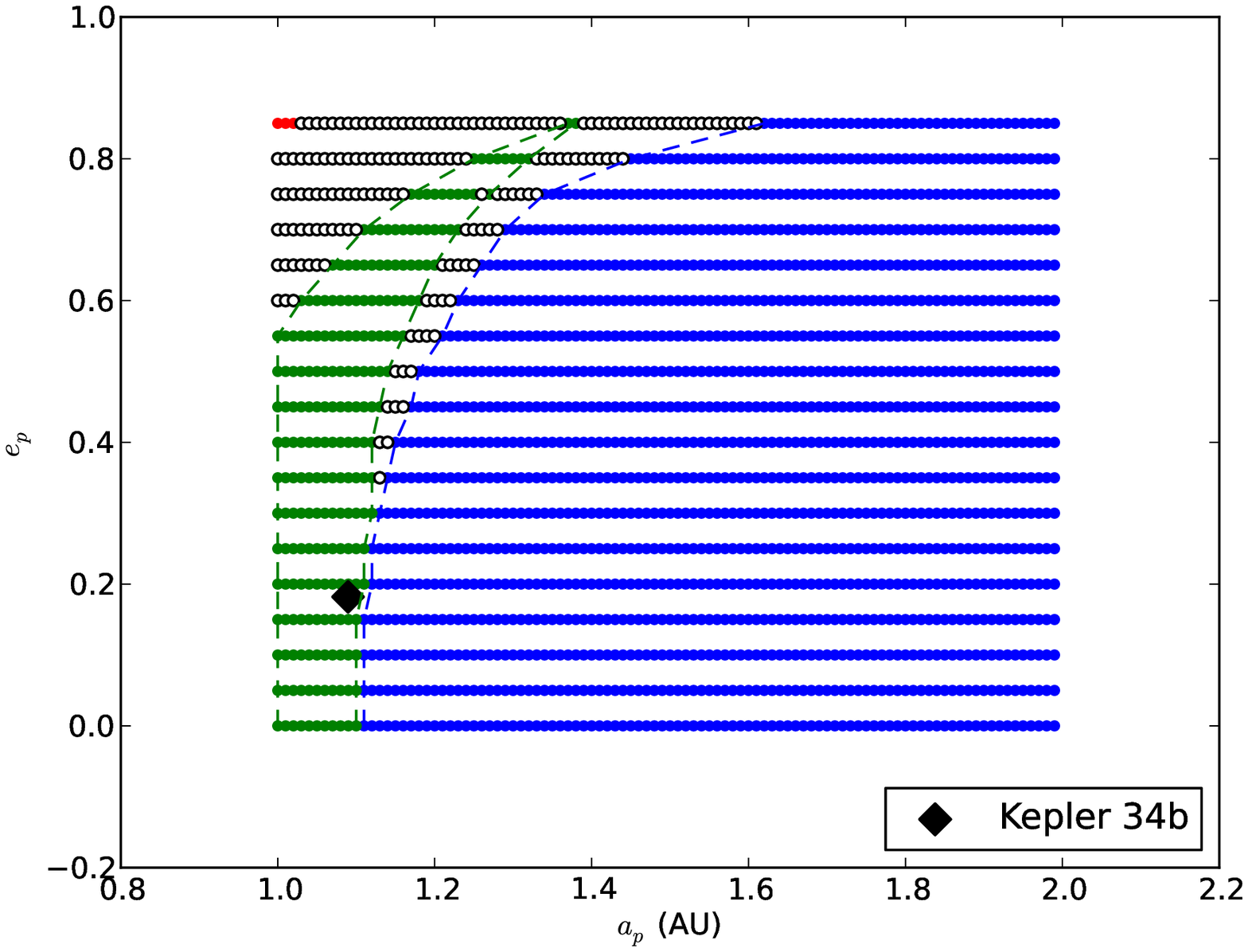} \\
\end{array}$
\caption{Left: The habitable zone for an Earth-like planet in the Kepler-34 binary system; Right: the habitable zone for an Earth-like planet in the Kepler-34 system with the secondary removed. The points are coloured according to the same classification system as previous figures. \label{fig:kepler34ab}}
\end{center}
\end{figure*}

\noindent Kepler-34 is a circumbinary planetary system possessing two G type stars, first reported in \citet{Welsh2012}, with stellar masses very close to equal, in a 28 day orbit.  The semi-major axis of the binary orbit is similar to that of Kepler-16, but the eccentricity is quite large.  Figure \ref{fig:kepler34ab} shows the habitable zones in the case where the secondary of Kepler-34 is either present (left panel) or absent (right panel).  The effect of adding the secondary to the system is significant, pushing the outer HZ boundary from around 1.1 AU at zero eccentricity to around 1.5 AU.  This shift is so extreme that there are very few simulations that are classified as habitable (green) in both cases (including the parameters corresponding to the planet Kepler-34b).  The inner and outer edges of the habitable zone meet at a peak at $e=0.9$ in the single star case - in the two star case, the height of this peak is reduced from $e=0.9$ to 0.8.

Kepler-34b has a semi-major axis of 1.0896 AU, with an eccentricity of 0.182 \citep{Welsh2012}.  The simulation corresponding most closely to these parameters is classified as transient, although this is somewhat moot given that the planet mass is 0.22 $\mjup$.  This being the case, it could still be a promising host for a habitable exomoon, as the cooling effect of eclipses of the moon by Kepler-34b itself may help to make the surface more clement \citep{Heller2012,Forgan_moon1}.  Ironically, if the secondary is removed from the simulation (right panel), Kepler-34b is very much inside the habitable zone.

\subsection{Kepler-35}

\begin{figure*}
\begin{center}$\begin{array}{cc}
\includegraphics[scale = 0.5]{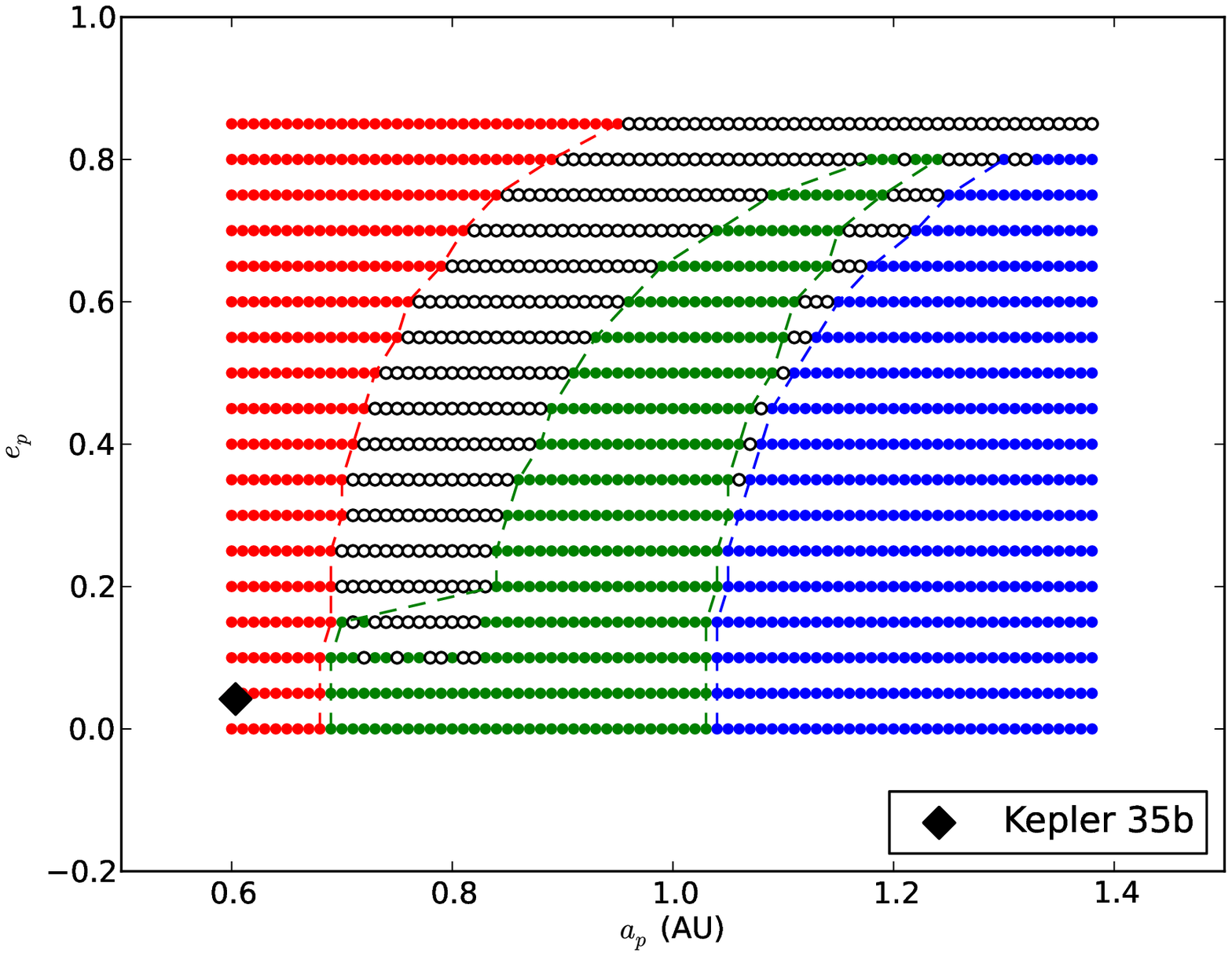} &
\includegraphics[scale=0.5]{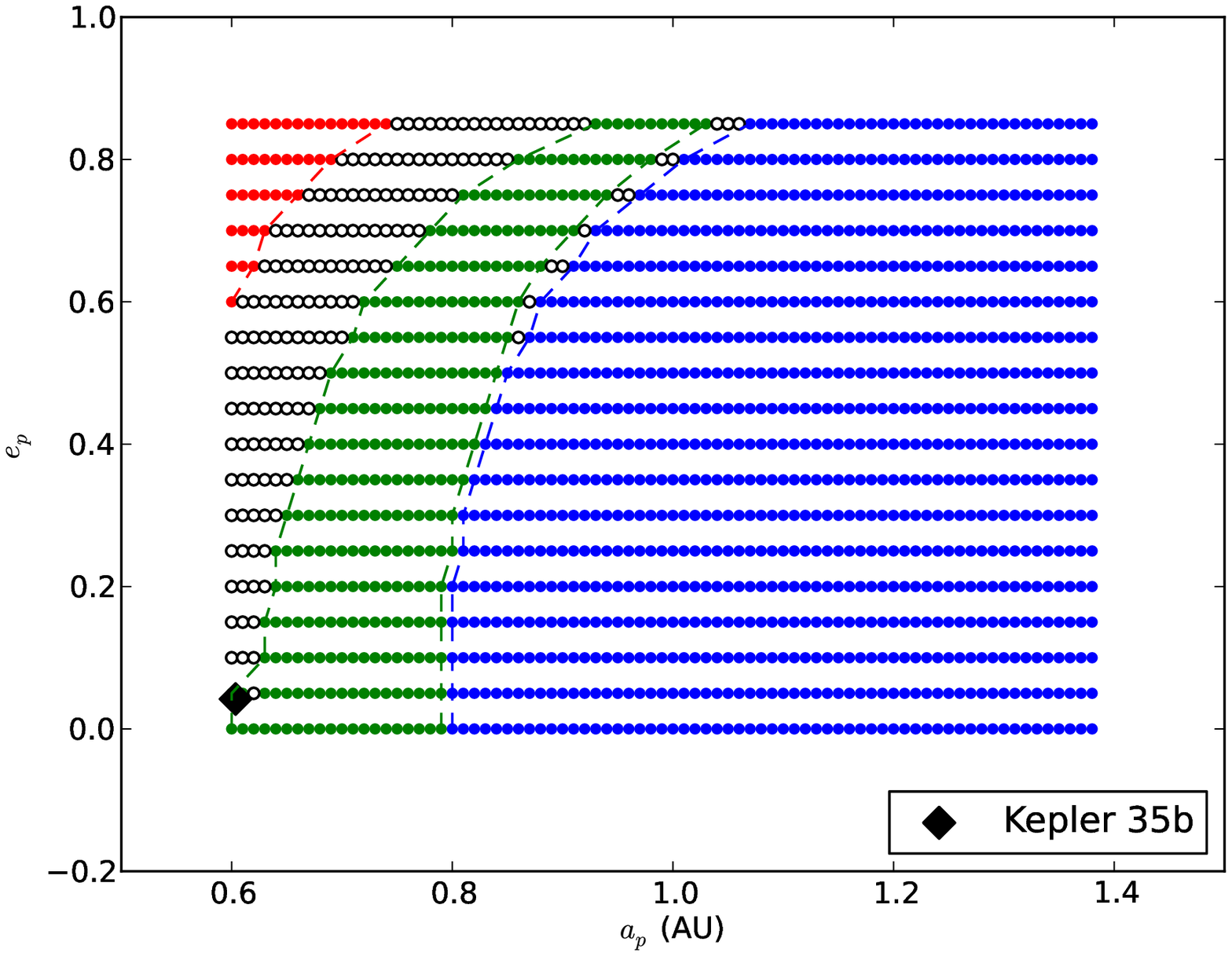} \\
\end{array}$
\caption{Left: The habitable zone for an Earth-like planet in the Kepler-35 binary system; Right: the habitable zone for an Earth-like planet in the Kepler-35 system with the secondary removed.  The points are coloured according to the same classification system as previous figures. \label{fig:kepler35ab}}
\end{center}
\end{figure*}

\noindent Reported alongside Kepler-34 in \citet{Welsh2012}, Kepler-35 also consists of two roughly equal mass G stars, but in a low eccentricity orbit with a period of 20 days.   Figure \ref{fig:kepler35ab} shows the habitable zones derived for this system (with and without the secondary star).  Again, as the binary masses are close to equal, the HZ boundaries shift significantly, and the highest eccentricity orbits are no longer continuously habitable.  

Kepler-35b orbits with a period of 131 days at a semi-major axis of 0.6 AU, in a low eccentricity orbit ($e_p=0.042$).  Our calculations indicate that Kepler-35b is too hot to be within the habitable zone, and it would seem unlikely that any moon it might possess would be habitable either.  Again, on removal of the secondary, the exoplanet would be in the habitable zone (in this case near the inner edge).

\subsection{Kepler-47}

\begin{figure*}
\begin{center}$\begin{array}{cc}
\includegraphics[scale = 0.5]{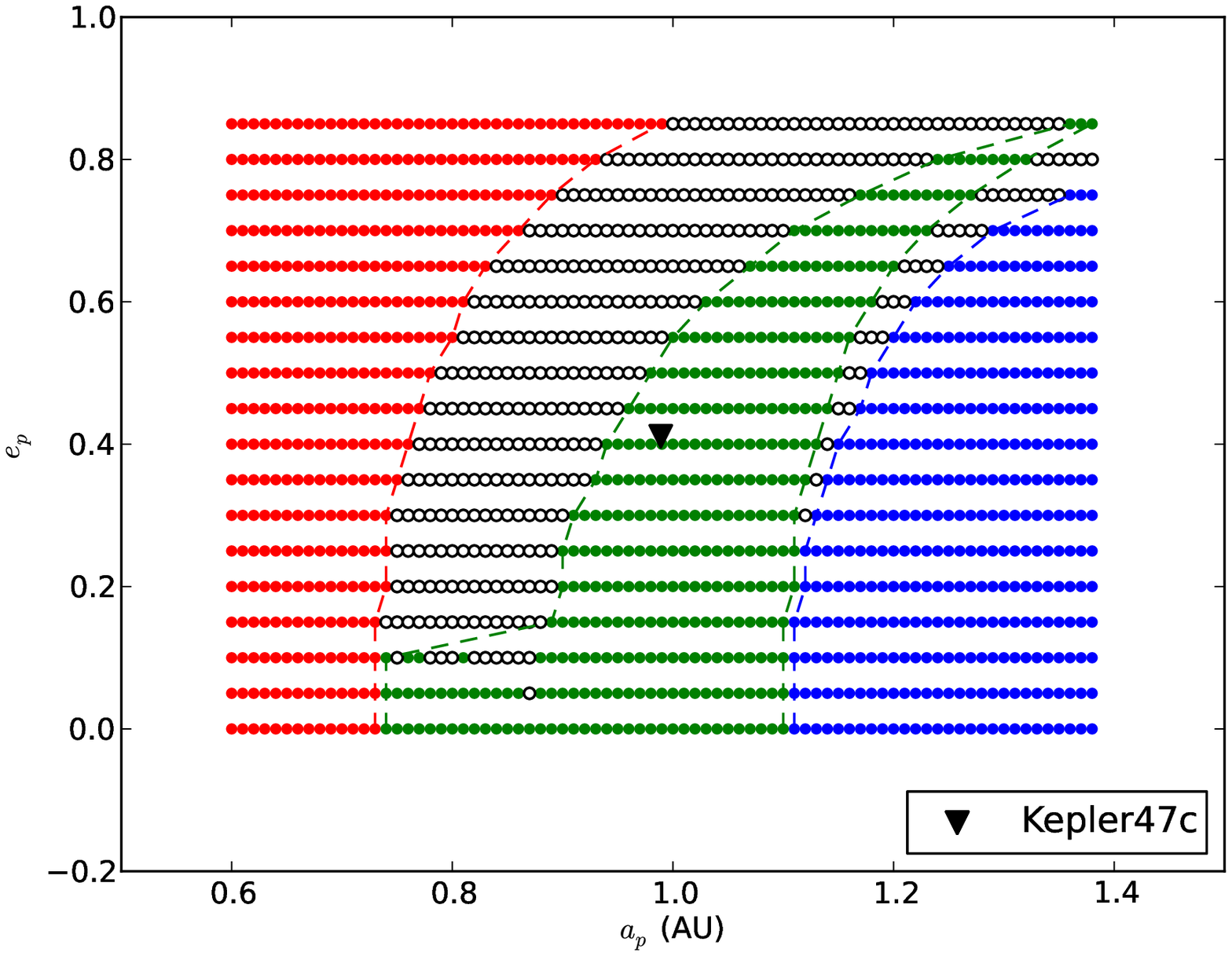} &
\includegraphics[scale=0.5]{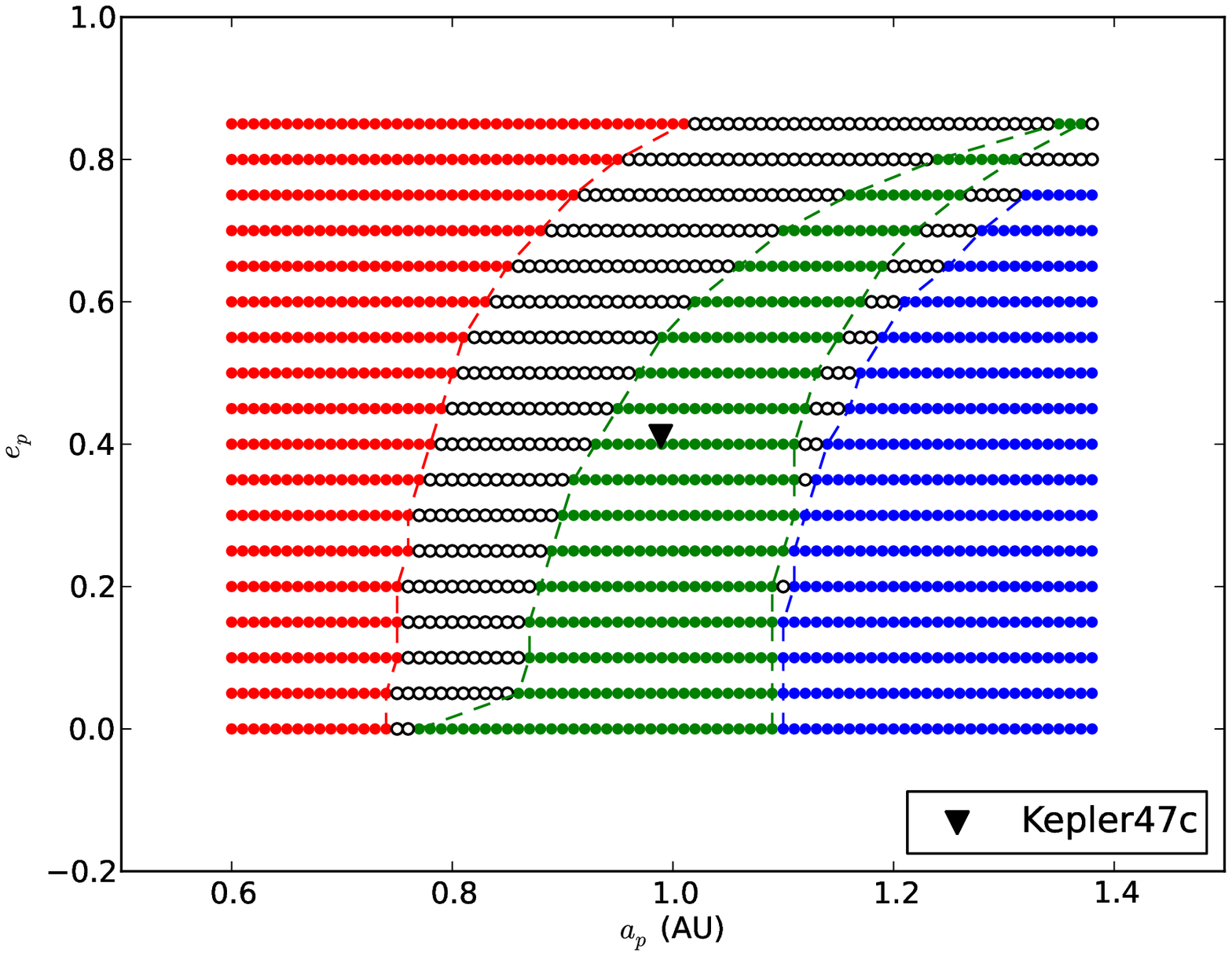} \\
\end{array}$
\caption{Left: The habitable zone for an Earth-like planet in the Kepler-47 binary system; Right: the habitable zone for an Earth-like planet in the Kepler-47 system with the secondary removed. The points are coloured according to the same classification system as previous figures. Kepler-47b is not marked as it is at lower semimajor axis than shown here.  Kepler-47c is marked as a downward triangle to represent that the eccentricity measured is an upper limit.\label{fig:kepler47ab}}
\end{center}
\end{figure*}

\noindent This binary system has the distinction of being the first P-type with multiple planets detected in orbit \citep{Orosz2012}.  Consisting of a G and M star in a tight low eccentricity orbit, the system has two planets orbiting in the binary plane at 49 days (Kepler-47b) and 303 days (Kepler-47c).  The outer planet is thought to be in the habitable zone, although with the eccentricity of the planet established only as an upper limit ($e_p < 0.411$), it is unclear how long it will spend in the spatial HZ, as noted by both \citet{Kane2013} and \citet{Haghighipour2013}.  Figure \ref{fig:kepler47ab} shows the orbital HZ constructed for the Kepler-47 binary system and for the Kepler-47 primary alone.  The presence of the M star has little effect on the HZ - the increased flux allows low eccentricity planets to be more habitable at semi-major axes between 0.7 and 0.9 AU, but otherwise there is little else to report. 

Kepler-47b has a near circular orbit at 0.2956 AU, and is clearly not in the habitable zone.  Despite its currently uncertain eccentricity, Kepler-47c does indeed appear to be warm and habitable, with low climate variations.  If Kepler-47c possessed an eccentricity larger than around 0.5, then it would fall into the region of parameter space occupied by transient classifications. Again, we should really only consider moons of Kepler-47c for Earthlike habitability, as the planet is Neptune-sized \citep{Orosz2012}.

\subsection{PH-1}

\begin{figure*}
\begin{center}$\begin{array}{cc}
\includegraphics[scale = 0.5]{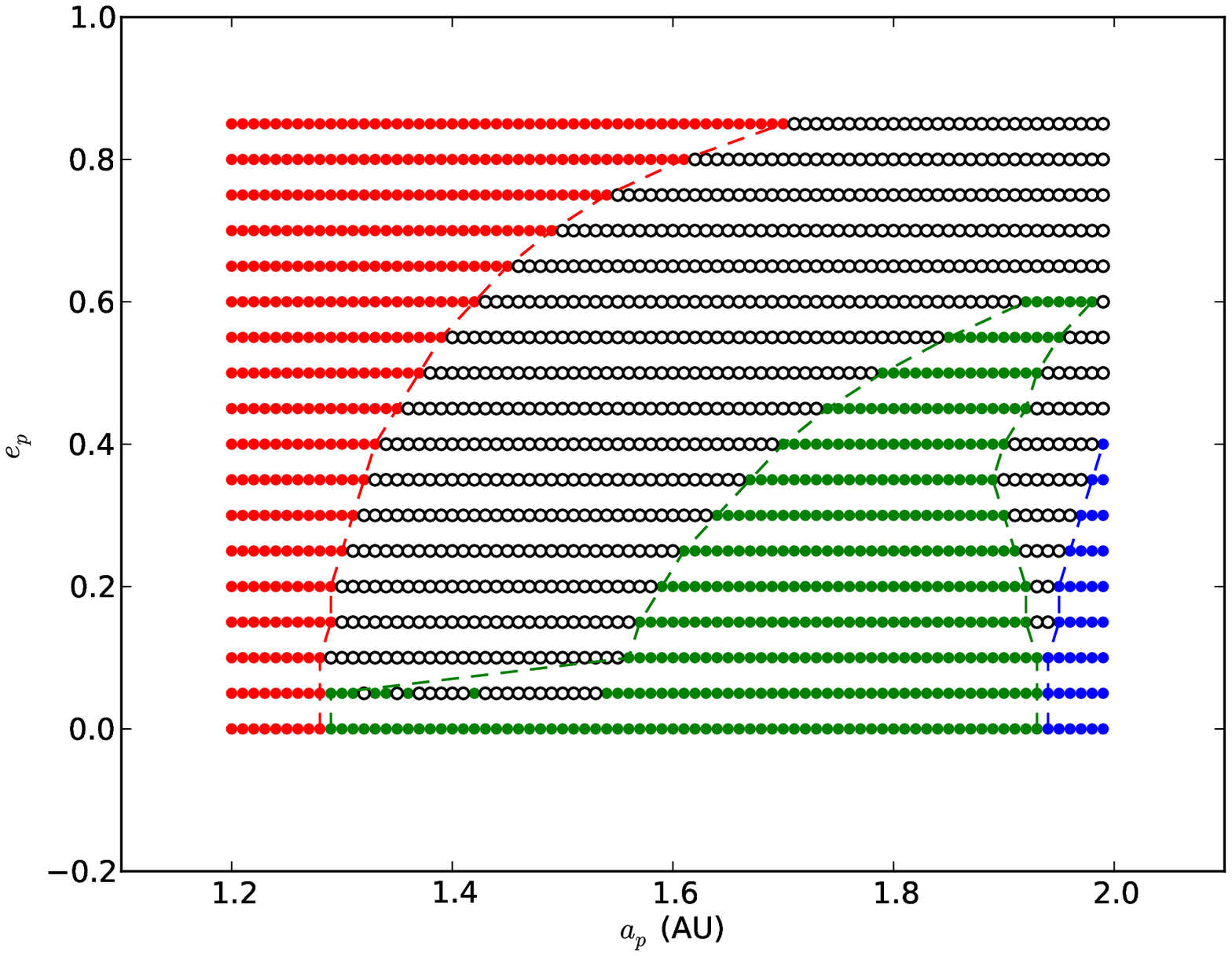}  &
\includegraphics[scale=0.5]{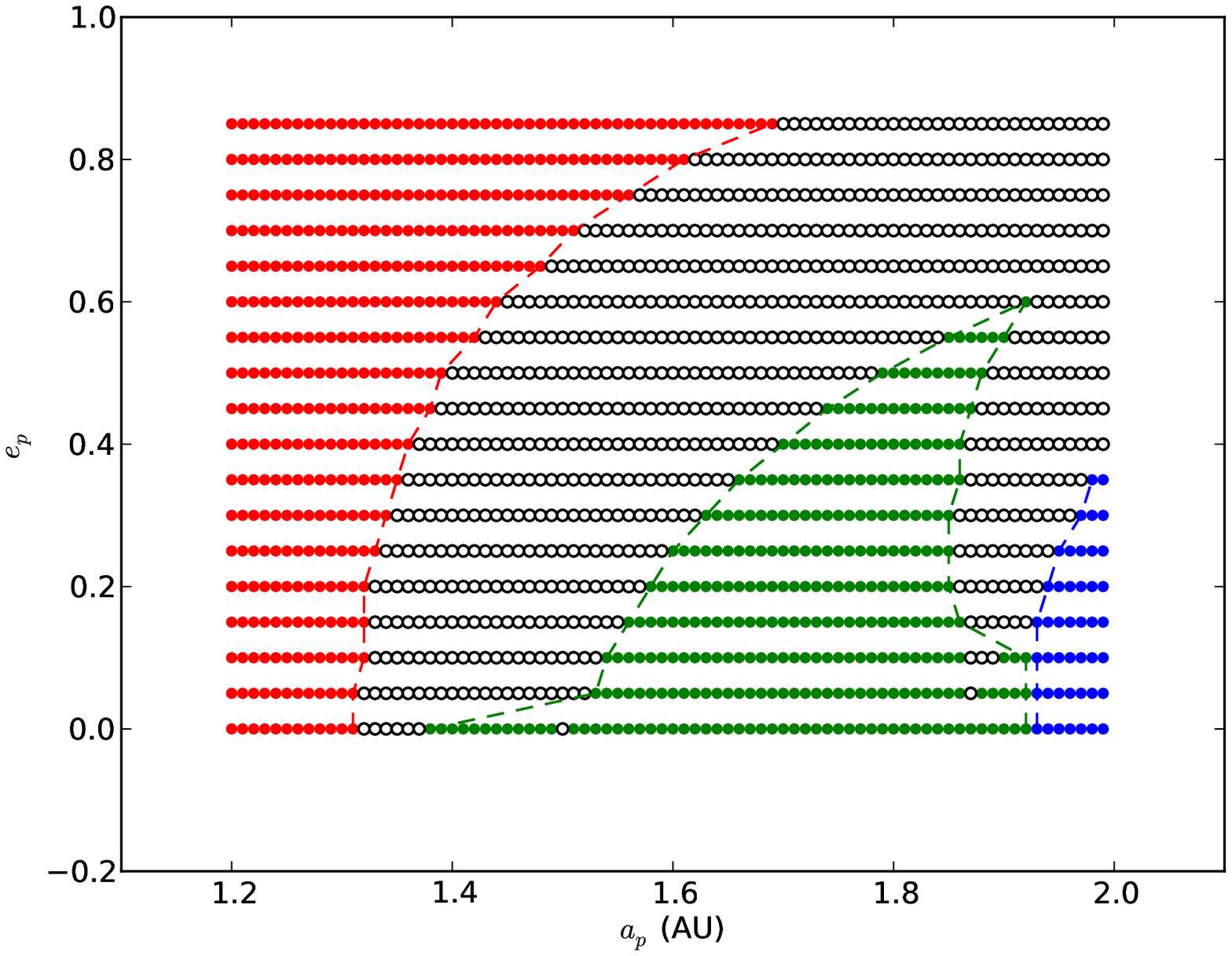} \\
\end{array}$
\caption{Left: The habitable zone for an Earth-like planet in the PH1 binary system; Right: the habitable zone for an Earth-like planet in the PH1 system with the secondary removed. The points are coloured according to the same classification system as previous figures.  PH1b is not plotted as its semimajor axis is less than 1.2 AU.\label{fig:ph1}}
\end{center}
\end{figure*}

\noindent PH-1 (also designated Kepler-64) is a quadruple star system, with the planet PH1b orbiting a F and M binary system.  The other two stars in the PH1 system form a separate binary which orbits at a distance of 1000 AU, which is sufficiently distant to neglect their fluxes.  The planet was detected by the PlanetHunters citizen science program \citep{Schwamb2013}, and orbits with a period of 138 days.  This system possesses the most extreme stellar mass ratio, and this is reflected in Figure \ref{fig:ph1}, which shows the orbital HZs constructed both with and without the presence of the M star.  The two figures are close to identical, with the exception of low eccentricity, low semimajor axis planets becoming slightly more habitable when the secondary is added, thanks to the cooling effect of eclipses.  In the single star case (right panel), the inner and outer HZ edges meet at $e_p=0.6$ - with the addition of the second star, the inner and outer edges no longer meet, as the outer edge is pushed to 1.99 AU.

The binary orbital period is 20 days (as the eccentricity is not constrained by observations, we assume the binary orbit is circular).  Hence, as habitable planets will orbit with periods of 500 days or more, the effect of such frequent eclipses on the planetary climate is softened by the atmospheric thermal inertia of the planet.  As a 0.5 $\mjup$ planet orbiting well within the inner HZ boundary, it is clear that PH1b is not habitable, and is unlikely to host habitable moons.

\section{Discussion}\label{sec:Discussion}

\subsection{Stability of Orbits in the Habitable Zone}

\noindent In this work, we do not model the gravitational force of the binary upon the planet.  Instead, we simply assume fixed Keplerian orbits for the planets around the barycentre of the binary system.  By doing so, we ignore the complicating factors of forming habitable planets with the orbital elements within our parameter study, either \emph{in situ} or through subsequent migration (cf \citealt{Meschiari2012} and \citealt{Dunhill2013}'s studies of Kepler-16b).  More importantly, we have not yet considered if these orbits are expected to be stable.

\citet{Holman1999a} simulate the motion of test particles in both P-type and S-type planetary systems, and produce fitting formulae for a critical semi-major axis for orbital stability.  In the case of P-type systems, this critical semi-major axis is a minimum, and is sensitive to the binary's orbital elements and the mass ratio of the binary $\mu = M_2/(M_1+M_2)$:

\begin{displaymath} 
a_p > a_{min} = a_{bin}\left(1.6 + 5.1 e_{bin} + 4.12 \mu + 2.22 e^2_{bin} \right.
\end{displaymath}

\begin{equation}
\left. -4.27 \mu e_{bin} -5.09 \mu^2 + 4.61 \mu^2 e^2_{bin}\right). \label{eq:stable_amin}
 \end{equation}

\noindent From these equations, it becomes clear that the frequency of planetary systems with potentially habitable planets will depend in the first instance on the Galactic binary fraction and their orbital statistics (see e.g. \citealt{Parker2013}).

Note that this stability prescription ignores the potential for instability islands at $a_p > a_{min}$ due to mean motion resonances, and if orbits out of the binary plane are permitted (which we have not considered here), then the dynamical landscape is quite rich (see e.g. \citealt{Doolin2011}).

%


Kepler-16 does not have a stable orbital HZ according to this analysis, whereas the other systems have HZs well clear of the instability limit.  We should note that the LEBM may underestimate the extent of the outer HZ boundary due to some missing physics (see section \ref{sec:limitations}), but it remains clear that low mass circumbinary systems are not promising candidates for stable habitable zones.

\subsection{Dependence on Binary Orbital Elements}

\noindent The P-type systems investigated here give a useful indication of how the habitable zone changes as the binary mass ratio $\mu$, binary semimajor axis $a_{bin}$ and binary eccentricity $e_{bin}$ are altered.  However, it is also instructive to select a single system and vary its parameters.  We use Kepler-35 for this analysis, as the previous shows that systems with $\mu \sim 0.5$ tend to produce orbital HZs more sensitive to the binary orbital elements (in line with previous spatial HZ analyses), and we wish to consider HZs that are still orbitally stable as $e_{bin}$ and $a_{bin}$ are changed.

\begin{figure*}
\begin{center}$\begin{array}{cc}
\includegraphics[scale = 0.5]{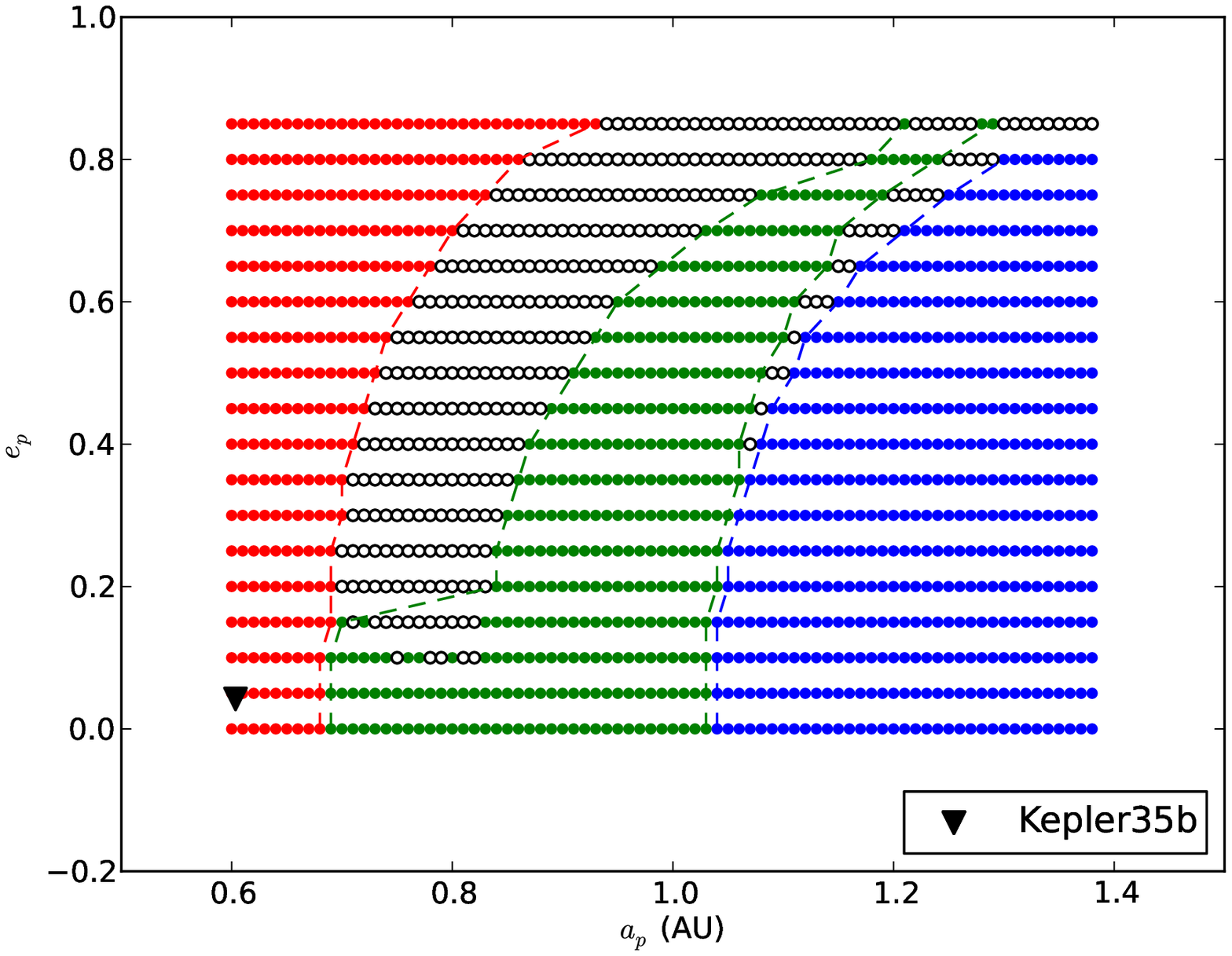} &
\includegraphics[scale = 0.5]{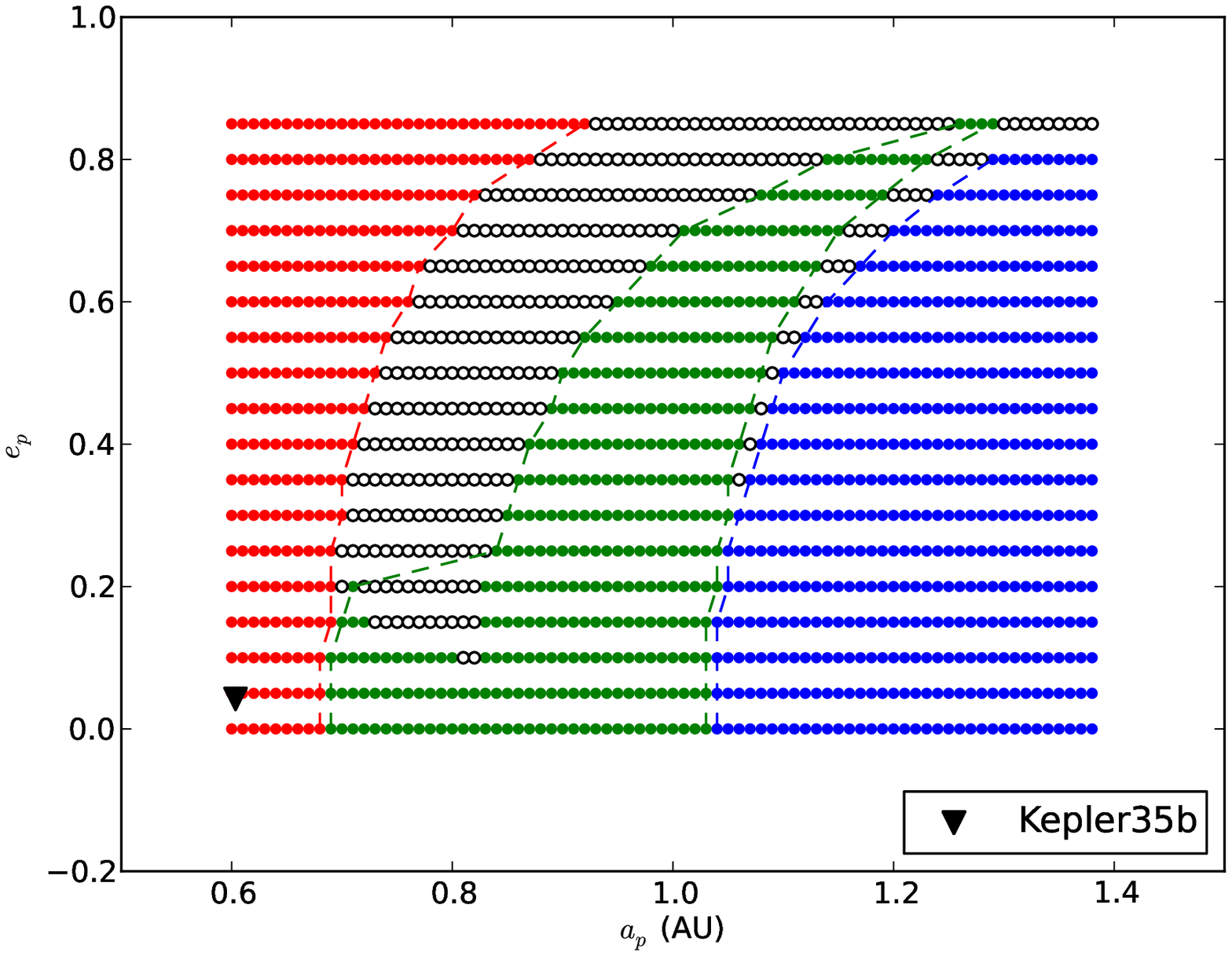} \\
\end{array}$
\caption{Left: The habitable zone for an Earth-like planet in the Kepler-35 binary system if the binary eccentricity is increased to $e_{bin}=0.3$.  Right: the same, with the binary eccentricity increased to $e_{bin}=0.5$. \label{fig:kepler35ab_highe}}
\end{center}
\end{figure*}

Figure \ref{fig:kepler35ab_highe} shows the Kepler-35 system orbital HZ as $e_{bin}$ is increased from 0.1418 to 0.3 (left panel) and 0.5 (right panel).  The minimum stable semimajor axes are $a_{min}= 0.567$ AU and 0.642 AU respectively.  The habitability of planets at low eccentricity and low semi-major axis appears to increase as a result of increasing $e_{bin}$, while remaining dynamically stable.  The inner and outer edges of the habitable become extended in eccentricity, all the way to $e_p=0.9$.  Objects with these orbital elements are likely to still possess strong temperature fluctuations on seasonal and dynamical timescales, but they appear to be just low enough to be less than 10\% of the mean temperature, and are therefore conventionally habitable.

\begin{figure*}
\begin{center}$\begin{array}{cc}
\includegraphics[scale = 0.5]{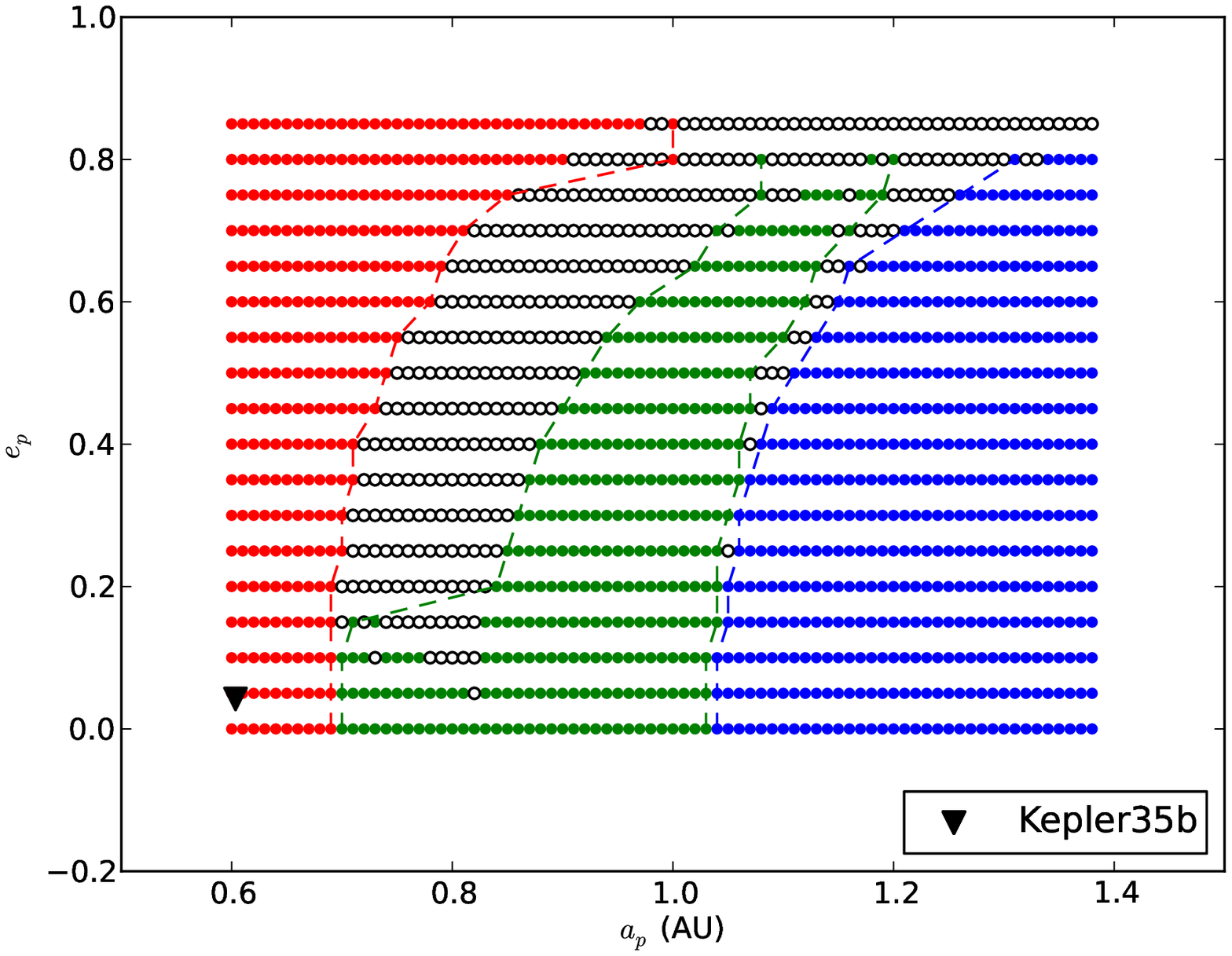} &
\includegraphics[scale = 0.5]{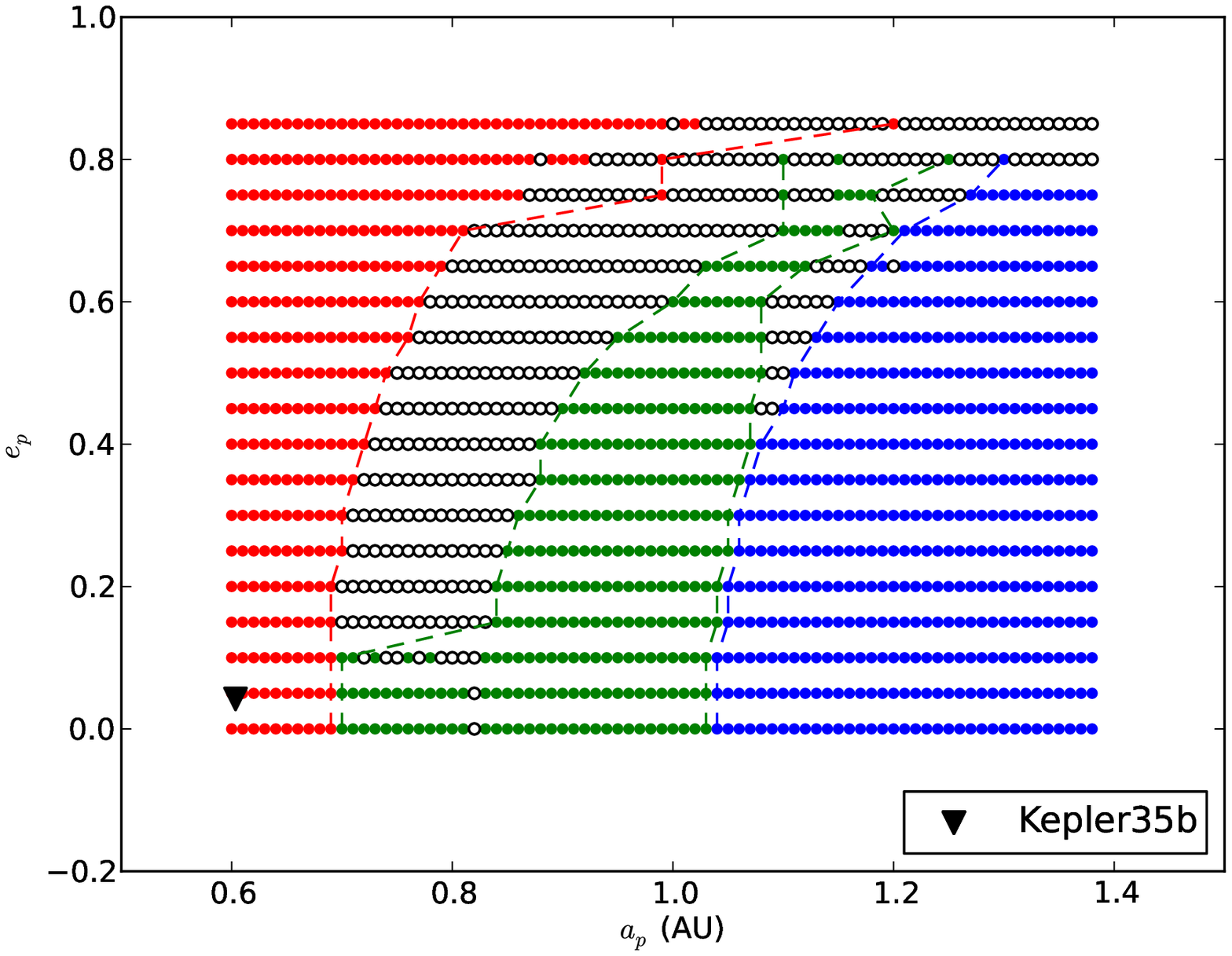} \\
\end{array}$
\caption{Left: The habitable zone for an Earth-like planet in the Kepler-35 binary system if the binary semimajor axis is increased to $a_{bin}=0.25$ AU. Right: the same, with the binary semimajor axis increased to $a_{bin}=0.3$ AU (right).  \label{fig:kepler35ab_higha}}
\end{center}
\end{figure*}

If we now return $e_{bin}$ to the standard value for Kepler-35 and increase $a_{bin}$ instead, we can see that the well-defined boundaries between habitable and non-habitable regions begin to blur.  An increase of $a_{bin}$ from 0.176 AU to 0.25 AU (left panel of Figure \ref{fig:kepler35ab_higha}) prevents planets with high eccentricity being habitable, and deepens a region of variable habitability at $a_p= 0.82$ AU, extending it downwards to $e_p=0.05$.  This corresponds to a planetary orbital period of 208 days, which is close to a 6:1 resonance with the binary (which now has an orbital period of 35 days).  The minimum stable orbit for this configuration is 0.705 AU, implying a small fraction of the inner HZ at low $a_p$, low $e_p$ cannot be considered habitable.

Increasing $a_{bin}$ to 0.3 AU (right panel of Figure \ref{fig:kepler35ab_higha}) serves only to exacerbate these issues.  The variable habitability at 0.82 AU now extends to circular planetary orbits, despite no longer corresponding to any orbital resonance.  The inner and outer boundaries of the HZ at high $e_p$ are no longer smooth, with the ability to determine whether a planet is continuously or variably habitable (green or clear) depending sensitively on the time period used to carry out the averaging.  The minimum stable orbit moves to 0.845 AU, rendering most of the inner HZ dynamically unstable. 

\subsection{Limitations of the Model \label{sec:limitations}}


\noindent Using a LEBM by definition requires some concessions to simplicity, especially if the goal is to run a large number of simulations.   However, we acknowledge that there are some potential improvements that should be considered in future work.

As previously mentioned, the orbits of the planets are fixed and Keplerian.  A more accurate representation would involve specifying a Keplerian orbit as initial conditions, and allowing the orbit to evolve under the gravitational influence of both stars in the system, either via full N Body calculations (e.g. \citealt{Meschiari2012}) or using analytic expressions which assume the planet mass to be negligible (e.g. \citealt{Leung2013}).  The non-Keplerian orbits that result from these calculations will add important variations in climate, which may make planets previously classified as ``habitable'' into ``transient'' planets (especially the low semi-major axis, low eccentricity planets that are commonly classified as habitable), or set up long term Milankovitch cycles \citep{Spiegel2010}.  

Also, the precession of periastron in circumbinary systems will strongly affect the range of seasonal variations eccentric planets will experience, as the longitude of solstices shifts further from the longitude of periastron \citep{Doolin2011,Armstrong2013}.  Kepler-34b is expected to undergo a complete cycle of periapse precession in around 20,000 years, whereas Kepler-35b is expected to do so in less than 10,000 years \citep{Welsh2012}.  These timescales are similar to the aforementioned Milankovitch cycles measured on Earth.  

When comparing this paper to analytical calculations of the spatial HZ, we find that our calculation of the outer HZ boundary (at zero eccentricity) is typically lower than that of the other authors.  This is most likely due to the rapid snowball albedo effect present due to the freezing of ice, which does not have a counter-opposing mechanism to suppress it (excluding adding more radiation to melt the ice).  In reality, more accurate modelling of the carbon-silicate cycle \citep{Williams1997a} would allow cooler planets to modify their atmospheric $CO_2$ levels.  As such, we do not fully model the ``maximum greenhouse'' conditions that are a standard of spatial HZ calculations, and this is an important feature that must be added to future models.

\section{Conclusions}\label{sec:Conclusions}

\noindent We have used one dimensional latitudinal energy balance modelling (LEBM) to investigate the habitable zones (HZs) of planets orbiting P-type star systems.  By running many models for each star system, an ``orbital HZ'' can be produced, which maps out the HZ in terms of the planet's orbital elements.  This numerical analysis is complementary to the common practice of mapping out the HZ in terms of its spatial extent using analytical calculations.   With the use of LEBMs, the orbital HZ allows the effects of stellar eclipses and planet eccentricity to be more simply incorporated.

We apply this technique to the circumbinary planetary systems Kepler-16, Kepler-34, Kepler-35, Kepler-47 and PH1.  In general, our orbital HZs are consistent with the spatial HZs derived by other authors (e.g. \citealt{Kane2013} and \citealt{Haghighipour2013}).  As has been found previously, the habitable zone strongly deviates from the single star HZ when the stars are of approximately equal mass.  If the primary is much more massive than the secondary, then the single star HZ and circumbinary HZs are very similar, although we note that in the case of Kepler-16, which contains two low mass stars with a relatively large mass difference, eclipses can become important.

Of the circumbinary planets orbiting the binaries we investigated, Kepler-47c was the only planet found to reside within the habitable zone.  We are able to make this determination despite the uncertainty of the planet's eccentricity, an advantage of orbital HZ modelling.  Kepler-47c is therefore an interesting target for future exomoon detections, as while Kepler-47c is not Earthlike, it may possess terrestrial moons.  Kepler-34b would be marginally habitable if it were of Earth mass - if it possesses an Earthlike moon, it may be able to sustain a biosphere, but it would need to be robust against strong oscillations in the moon's climate.

While we have not explicitly simulated the orbital stability of these planets, previous analytical calculations of the minimum semimajor axis for a stable circumbinary orbit indicate that with the exception of Kepler-16b, the HZs produced in this work should be amenable to terrestrial planets on stable orbits.  However, the dynamical complexity of circumbinary systems warrants further investigation with more appropriate gravitational physics included. 

\section*{Acknowledgments}

\noindent DF gratefully acknowledges support from STFC grant ST/J001422/1.  The author would like to thank the referee for their invaluable comments which greatly improved this paper.

\bibliographystyle{mn2e} 
\bibliography{circumbinary_EBM}

\begin{thebibliography}{49}
\expandafter\ifx\csname natexlab\endcsname\relax\def\natexlab#1{#1}\fi

\bibitem[{Abe {et~al}\mbox{.}(2011)Abe, Abe-Ouchi, Sleep, \& Zahnle}]{Abe2011}
Abe Y., Abe-Ouchi A., Sleep N.~H., Zahnle K.~J., 2011, Astrobiology, 11, 443

\bibitem[{Anglada-Escud\'{e} {et~al}\mbox{.}(2013)Anglada-Escud\'{e}, Tuomi,
  Gerlach, Barnes, Heller, Jenkins, Wende, Vogt, {Paul Butler}, Reiners, \&
  Jones}]{Anglada-Escude2013}
Anglada-Escud\'{e} G. {et~al.}, 2013, A\&A, 556, A126

\bibitem[{Armstrong {et~al}\mbox{.}(2013)Armstrong, Martin, Brown, Faedi,
  {Gomez Maqueo Chew}, Mardling, Pollacco, Triaud, \& Udry}]{Armstrong2013}
Armstrong D. {et~al.}, 2013, MNRAS, in press

\bibitem[{Borucki {et~al}\mbox{.}(2013)Borucki, Agol, Fressin, Kaltenegger,
  Rowe, Isaacson, Fischer, Batalha, Lissauer, Marcy, Fabrycky, D\'{e}sert,
  Bryson, Barclay, Bastien, Boss, Brugamyer, Buchhave, Burke, Caldwell, Carter,
  Charbonneau, Crepp, Christensen-Dalsgaard, Christiansen, Ciardi, Cochran,
  DeVore, Doyle, Dupree, Endl, Everett, Ford, Fortney, Gautier, Geary, Gould,
  Haas, Henze, Howard, Howell, Huber, Jenkins, Kjeldsen, Kolbl, Kolodziejczak,
  Latham, Lee, Lopez, Mullally, Orosz, Prsa, Quintana, Sanchis-Ojeda, Sasselov,
  Seader, Shporer, Steffen, Still, Tenenbaum, Thompson, Torres, Twicken, Welsh,
  \& Winn}]{Borucki2013}
Borucki W.~J. {et~al.}, 2013, Science (New York, N.Y.), 340, 587

\bibitem[{Borucki {et~al}\mbox{.}(2012)Borucki, Koch, Batalha, Bryson, Rowe,
  Fressin, Torres, Caldwell, Christensen-Dalsgaard, Cochran, DeVore, Gautier,
  Geary, Gilliland, Gould, Howell, Jenkins, Latham, Lissauer, Marcy, Sasselov,
  Boss, Charbonneau, Ciardi, Kaltenegger, Doyle, Dupree, Ford, Fortney, Holman,
  Steffen, Mullally, Still, Tarter, Ballard, Buchhave, Carter, Christiansen,
  Demory, D\'{e}sert, Dressing, Endl, Fabrycky, Fischer, Haas, Henze, Horch,
  Howard, Isaacson, Kjeldsen, Johnson, Klaus, Kolodziejczak, Barclay, Li,
  Meibom, Prsa, Quinn, Quintana, Robertson, Sherry, Shporer, Tenenbaum,
  Thompson, Twicken, {Van Cleve}, Welsh, Basu, Chaplin, Miglio, Kawaler,
  Arentoft, Stello, Metcalfe, Verner, Karoff, Lundkvist, Lund, Handberg,
  Elsworth, Hekker, Huber, Bedding, \& Rapin}]{Borucki2012c}
Borucki W.~J. {et~al.}, 2012, ApJ, 745, 120

\bibitem[{{Del Genio}(1993)}]{DelGenio1993}
{Del Genio} A., 1993, Icarus, 101, 1

\bibitem[{{Del Genio}(1996)}]{DelGenio1996}
{Del Genio} A., 1996, Icarus, 120, 332

\bibitem[{Doolin \& Blundell(2011)}]{Doolin2011}
Doolin S., Blundell K.~M., 2011, MNRAS, 418, 2656

\bibitem[{Doyle {et~al}\mbox{.}(2011)Doyle, Carter, Fabrycky, Slawson, Howell,
  Winn, Orosz, Pr\v{s}a, Welsh, Quinn, Latham, Torres, Buchhave, Marcy,
  Fortney, Shporer, Ford, Lissauer, Ragozzine, Rucker, Batalha, Jenkins,
  Borucki, Koch, Middour, Hall, McCauliff, Fanelli, Quintana, Holman, Caldwell,
  Still, Stefanik, Brown, Esquerdo, Tang, Furesz, Geary, Berlind, Calkins,
  Short, Steffen, Sasselov, Dunham, Cochran, Boss, Haas, Buzasi, \&
  Fischer}]{Doyle2011}
Doyle L.~R. {et~al.}, 2011, Science (New York, N.Y.), 333, 1602

\bibitem[{Dressing {et~al}\mbox{.}(2010)Dressing, Spiegel, Scharf, Menou, \&
  Raymond}]{Dressing2010}
Dressing C.~D., Spiegel D.~S., Scharf C.~A., Menou K., Raymond S.~N., 2010,
  ApJ, 721, 1295

\bibitem[{Dunhill \& Alexander(2013)}]{Dunhill2013}
Dunhill A., Alexander R., 2013, MNRAS, in press

\bibitem[{Eggl {et~al}\mbox{.}(2012)Eggl, Pilat-Lohinger, Georgakarakos,
  Gyergyovits, \& Funk}]{Eggl2012}
Eggl S., Pilat-Lohinger E., Georgakarakos N., Gyergyovits M., Funk B., 2012,
  ApJ, 752, 74

\bibitem[{Farrell(1990)}]{Farrell1990}
Farrell B.~F., 1990, Journal of Atmospheric Sciences, 47, 2986

\bibitem[{Forgan(2012)}]{Forgan2012}
Forgan D., 2012, MNRAS, 422, 1241

\bibitem[{Forgan \& Kipping(2013)}]{Forgan_moon1}
Forgan D., Kipping D., 2013, MNRAS, 432, 2994

\bibitem[{Haghighipour \& Kaltenegger(2013)}]{Haghighipour2013}
Haghighipour N., Kaltenegger L., 2013, ApJ, in press

\bibitem[{Hart(1979)}]{Hart_HZ}
Hart M.~H., 1979, Icarus, 37, 351

\bibitem[{Heller(2012)}]{Heller2012}
Heller R., 2012, A\&A, 545, L8

\bibitem[{Holman \& Wiegert(1999)}]{Holman1999a}
Holman M.~J., Wiegert P.~A., 1999, The Astronomical Journal, 117, 621

\bibitem[{Huang(1959)}]{Huang1959}
Huang S.-S., 1959, PASP, 71, 421

\bibitem[{Kaltenegger \& Haghighipour(2013)}]{Kaltenegger2013}
Kaltenegger L., Haghighipour N., 2013, ApJ, in press

\bibitem[{Kaltenegger \& Sasselov(2011)}]{Kaltenegger2011}
Kaltenegger L., Sasselov D., 2011, ApJ, 736, L25

\bibitem[{Kane \& Gelino(2012{\natexlab{a}})}]{Kane2012}
Kane S.~R., Gelino D.~M., 2012{\natexlab{a}}, Astrobiology, 12, 940

\bibitem[{Kane \& Gelino(2012{\natexlab{b}})}]{Kane2012c}
Kane S.~R., Gelino D.~M., 2012{\natexlab{b}}, PASP, 124, 323

\bibitem[{Kane \& Hinkel(2013)}]{Kane2013}
Kane S.~R., Hinkel N.~R., 2013, ApJ, 762, 7

\bibitem[{Kasting, Whitmire \& Reynolds(1993)Kasting, Whitmire, \&
  Reynolds}]{Kasting_et_al_93}
Kasting J., Whitmire D., Reynolds R., 1993, Icarus, 101, 108

\bibitem[{Kipping {et~al}\mbox{.}(2013)Kipping, Forgan, Hartman, Nesvorny,
  Bakos, Schmitt, \& Buchhave}]{Kipping2013a}
Kipping D.~M., Forgan D., Hartman J., Nesvorny D., Bakos G.~A., Schmitt A.~R.,
  Buchhave L.~A., 2013, ApJ, in press

\bibitem[{Kopparapu {et~al}\mbox{.}(2013)Kopparapu, Ramirez, Kasting, Eymet,
  Robinson, Mahadevan, Terrien, Domagal-Goldman, Meadows, \&
  Deshpande}]{Kopparapu2013}
Kopparapu R.~K. {et~al.}, 2013, ApJ, 765, 131

\bibitem[{Leung \& Lee(2013)}]{Leung2013}
Leung G. C.~K., Lee M.~H., 2013, ApJ, 763, 107

\bibitem[{Mayor \& Queloz(1995)}]{Mayor1995}
Mayor M., Queloz D., 1995, Nature, 378, 355

\bibitem[{Meschiari(2012)}]{Meschiari2012}
Meschiari S., 2012, ApJ, 752, 71

\bibitem[{North, Cahalan \& Coakley(1981)North, Cahalan, \&
  Coakley}]{North1981}
North G., Cahalan R., Coakley J., 1981, Rev. Geophys. Space Phys., 19, 91

\bibitem[{Orosz {et~al}\mbox{.}(2012)Orosz, Welsh, Carter, Fabrycky, Cochran,
  Endl, Ford, Haghighipour, MacQueen, Mazeh, Sanchis-Ojeda, Short, Torres,
  Agol, Buchhave, Doyle, Isaacson, Lissauer, Marcy, Shporer, Windmiller,
  Barclay, Boss, Clarke, Fortney, Geary, Holman, Huber, Jenkins, Kinemuchi,
  Kruse, Ragozzine, Sasselov, Still, Tenenbaum, Uddin, Winn, Koch, \&
  Borucki}]{Orosz2012}
Orosz J.~A. {et~al.}, 2012, Science (New York, N.Y.), 337, 1511

\bibitem[{Parker \& Quanz(2013)}]{Parker2013}
Parker R.~J., Quanz S.~P., 2013, MNRAS, in press

\bibitem[{Quarles, Musielak \& Cuntz(2012)Quarles, Musielak, \&
  Cuntz}]{Quarles2012}
Quarles B., Musielak Z.~E., Cuntz M., 2012, ApJ, 750, 14

\bibitem[{Quintana {et~al}\mbox{.}(2007)Quintana, Adams, Lissauer, \&
  Chambers}]{Quintana2007}
Quintana E.~V., Adams F.~C., Lissauer J.~J., Chambers J.~E., 2007, ApJ, 660,
  807

\bibitem[{Quintana {et~al}\mbox{.}(2002)Quintana, Lissauer, Chambers, \&
  Duncan}]{Quintana2002}
Quintana E.~V., Lissauer J.~J., Chambers J.~E., Duncan M.~J., 2002, ApJ, 576,
  982

\bibitem[{Schwamb {et~al}\mbox{.}(2013)Schwamb, Orosz, Carter, Welsh, Fischer,
  Torres, Howard, Crepp, Keel, Lintott, Kaib, Terrell, Gagliano, Jek, Parrish,
  Smith, Lynn, Simpson, Giguere, \& Schawinski}]{Schwamb2013}
Schwamb M.~E. {et~al.}, 2013, ApJ, 768, 127

\bibitem[{Selsis {et~al}\mbox{.}(2007)Selsis, Kasting, Levrard, Paillet, Ribas,
  \& Delfosse}]{Selsis2007}
Selsis F., Kasting J.~F., Levrard B., Paillet J., Ribas I., Delfosse X., 2007,
  A\&A, 476, 1373

\bibitem[{Spiegel, Menou \& Scharf(2008)Spiegel, Menou, \&
  Scharf}]{Spiegel_et_al_08}
Spiegel D.~S., Menou K., Scharf C.~A., 2008, ApJ, 681, 1609

\bibitem[{Spiegel, Menou \& Scharf(2009)Spiegel, Menou, \&
  Scharf}]{Spiegel2009}
Spiegel D.~S., Menou K., Scharf C.~A., 2009, ApJ, 691, 596

\bibitem[{Spiegel {et~al}\mbox{.}(2010)Spiegel, Raymond, Dressing, Scharf, \&
  Mitchell}]{Spiegel2010}
Spiegel D.~S., Raymond S.~N., Dressing C.~D., Scharf C.~A., Mitchell J.~L.,
  2010, ApJ, 721, 1308

\bibitem[{Spiegel \& Turner(2012)}]{Spiegel2012}
Spiegel D.~S., Turner E.~L., 2012, Proceedings of the National Academy of
  Sciences of the United States of America, 109, 395

\bibitem[{Underwood, Jones \& Sleep(2003)Underwood, Jones, \&
  Sleep}]{Underwood2003}
Underwood D., Jones B., Sleep P., 2003, International Journal of Astrobiology,
  2, 289

\bibitem[{Vladilo {et~al}\mbox{.}(2013)Vladilo, Murante, Silva, Provenzale,
  Ferri, \& Ragazzini}]{Vladilo2013}
Vladilo G., Murante G., Silva L., Provenzale A., Ferri G., Ragazzini G., 2013,
  ApJ, 767, 65

\bibitem[{Welsh {et~al}\mbox{.}(2012)Welsh, Orosz, Carter, Fabrycky, Ford,
  Lissauer, Pr\v{s}a, Quinn, Ragozzine, Short, Torres, Winn, Doyle, Barclay,
  Batalha, Bloemen, Brugamyer, Buchhave, Caldwell, Caldwell, Christiansen,
  Ciardi, Cochran, Endl, Fortney, Gautier, Gilliland, Haas, Hall, Holman,
  Howard, Howell, Isaacson, Jenkins, Klaus, Latham, Li, Marcy, Mazeh, Quintana,
  Robertson, Shporer, Steffen, Windmiller, Koch, \& Borucki}]{Welsh2012}
Welsh W.~F. {et~al.}, 2012, Nature, 481, 475

\bibitem[{Wiegert \& Holman(1997)}]{Wiegert1997}
Wiegert P.~A., Holman M.~J., 1997, The Astronomical Journal, 113, 1445

\bibitem[{Williams \& Kasting(1997)}]{Williams1997a}
Williams D., Kasting J.~F., 1997, Icarus, 129, 254

\bibitem[{Williams \& Pollard(2002)}]{Williams2002}
Williams D.~M., Pollard D., 2002, International Journal of Astrobiology, 1, 61

\end{thebibliography}

\appendix

\label{lastpage}

\end{document}